\documentclass[10pt, twoside, journal]{IEEEtran}

\usepackage{graphics}
\usepackage{amsmath}
\usepackage{latexsym}
\usepackage{amsbsy}
\usepackage{epstopdf}

\def\bfxi{\boldsymbol{\xi}}

\def\sce{\boldsymbol{\cal E}}
\def\sch{\boldsymbol{\cal H}}

\def\bfalpha{\boldsymbol{\alpha}}

\def\bfxi{\mbox{\boldmath$\xi$}}

\def\sce{{\mbox{\boldmath$\cal{E}$}}}
\def\sch{{\mbox{\boldmath$\cal{H}$}}}
\def\bfalpha{\mbox{\boldmath$\alpha$}}

\def\rmA{{\rm A}}
\def\rmB{{\rm B}}

\DeclareFontFamily{U}{cmbsy}{}
\DeclareFontShape{U}{cmbsy}{m}{n}{ <5> <6> <7> <8> <9> gen * cmbsy
       <10> <10.95> <12> <14.4> <17.28> <20.74> <24.88> cmbsy10}{}
\DeclareMathAlphabet{\calb}{U}{cmbsy}{m}{n}

\newcommand{\dyadic}[1]{{#1}
\setbox0=\hbox{$\scriptstyle\leftrightarrow$}
   \setbox2=\hbox{$#1$}
   \dimen0=.5\wd0 \advance\dimen0 by-.5\wd2
   \advance\dimen0 by-\wd0
   \kern\dimen0
{^{\hbox{$\scriptstyle\leftrightarrow$}}}}

\newcommand{\dyadictall}[1]{{#1}
\setbox0=\hbox{$\scriptstyle\leftrightarrow$}
   \setbox2=\hbox{$#1$}
   \dimen0=.5\wd0 \advance\dimen0 by-.5\wd2
   \advance\dimen0 by-\wd0
   \kern\dimen0
{^{\raise.5ex\hbox{$\scriptstyle\leftrightarrow$}}}}

\newcommand{\dyadictalll}[1]{{#1}
\setbox0=\hbox{$\scriptstyle\leftrightarrow$}
   \setbox2=\hbox{$#1$}
   \dimen0=.5\wd0 \advance\dimen0 by-.5\wd2
   \advance\dimen0 by-\wd0
   \kern\dimen0
{^{\raise.4ex\hbox{$\scriptstyle\leftrightarrow$}}}}

\begin{document}


\title{A Homogenization Technique for Obtaining Generalized Sheet Transition Conditions (GSTCs)
for a Metafilm Embedded in a Magneto-Dielectric Interface}

\author{~Christopher~L.~Holloway,~\IEEEmembership{Fellow,~IEEE,}
        Edward~F.~Kuester,~\IEEEmembership{Fellow,~IEEE}
        \thanks{Submitted to Journal Sept 2014: Revision is under review.}\thanks{C.L. Holloway, is with the National
Institute of Standards and Technology (NIST), Electromagnetics Division,
U.S. Department of Commerce, Boulder Laboratories,
Boulder,~CO~80305. E.F. Kuester is with the Department of Electrical, Computer and Energy Engineering, University of Colorado, Boulder, CO 80309. Publication of the U.S. government, not subject to U.S. copyright.}}

\markboth{Submitted to Journal Sept 2014: Revision is under review}{GSTCs for Metafilms}

\maketitle

\begin{abstract}

Using the multiple-scale homogenization method, we derive generalized sheet transition conditions (GSTCs) for electromagnetic fields at the surface of a metafilm. The scatterers that compose the metafilm are of arbitrary shape and are embedded between two different magneto-dielectric media. The parameters in these boundary conditions are interpreted as effective electric and magnetic surface susceptibilities, which themselves are related to the geometry of the scatterers that constitute the metafilm.
\vspace{7mm}

\end{abstract}

\section{Introduction}

\begin{figure}
\centering
\scalebox{0.55} {\includegraphics*{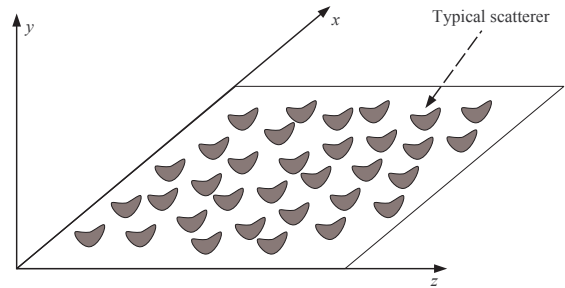}}
\caption{Illustration of a metafilm consisting of arbitrarily
shaped scatterers placed on the $x-z$ plane.}
\label{fig1}
\end{figure}

In this paper, we consider the interaction of electromagnetic waves with a two-dimensional periodic array of arbitrarily shaped
scatterers partially embedded between two different magneto-dielectric media, as shown in Fig. \ref{fig1}. This type of surface has been given the name metafilm \cite{kmh}, by which we specifically mean a surface distribution of separated electrically
small scatterers. As far as macroscopic fields are concerned, the metafilm acts as an infinitesimal
sheet---one that causes a phase shift and/or a change in amplitude in the fields
interacting with it. Scattering by such sheets is best characterized by generalized sheet-transition
conditions if computationally expensive numerical modeling is to be avoided \cite{kmh}.

There is currently a great deal of attention being focused on electromagnetic
metamaterials \cite{c1}-\cite{cui}---novel synthetic materials engineered to achieve unique properties not normally found in nature. Those unique properties promise a wide range of potential applications in the electromagnetic (EM) frequency ranges from RF to optical frequencies. Metamaterials are often engineered by arranging a set of scatterers throughout a region of space in a specific pattern so as to achieve some desirable bulk behavior of the material. This concept can be extended by judiciously placing scatterers in a two-dimensional pattern at a surface or interface. Such a surface version of a metamaterial has been given the name metasurface, and includes metafilms and metascreens as special cases \cite{hk3}, \cite{alex}. Metasurfaces have also been referred to in the literature as single-layer metamaterials.

The simplicity and relative ease of fabrication of metasurfaces makes them attractive alternatives to three-dimensional (3D) metamaterials; in many applications metasurfaces can be used in place of metamaterials. Metasurfaces have the advantage of taking up less physical space than do full 3D metamaterial structures; as a consequence they can also offer the possibility of lower losses. The application of metasurfaces at frequencies from microwave to optical has attracted great interest in recent years \cite{hk3}, \cite{alex}.

We will call any periodic two-dimensional structure whose thickness and periodicity are small compared to a wavelength in the surrounding media a metasurface. The distinction between a metasurface and a frequency-selective surface (FSS) is discussed in detail in \cite{hk3}.  Within this general designation, we can identify two important subclasses \cite{surfacewave}. Metasurfaces that have a ``cermet'' topology, which refers to an array of isolated (non-touching) scatterers are called metafilms, a term coined in \cite{kmh} for such surfaces. Metasurfaces with a ``fishnet'' structure are called metascreens \cite{hk3}. These are characterized by periodically spaced apertures in an otherwise relatively impenetrable surface. Other kinds of metasurfaces exist that lie somewhere between these two extremes. For example, a grating of parallel conducting wires behaves like a metafilm to electric fields perpendicular to the wire axes, but like a metascreen for electric fields parallel to the wire axes \cite{wirehk}. In this paper we will limit ourselves to metafilms. It is important to
note that the individual scatterers constituting the metafilm are not necessarily of zero thickness (or even small compared to the lattice constants); they may be of arbitrary shape, and their dimensions are required to be small only in comparison to a wavelength in the surrounding medium, \emph{a fortiori} because the lattice constant has been assumed small compared to a wavelength.

Like that of a metamaterial, the behavior of a metafilm can be understood in terms of the electric and magnetic polarizabilities of its constituent scatterers. The traditional and most convenient method by which to model metamaterials is with effective-medium theory, using the bulk electromagnetic parameters $\mu_{\rm eff}$ and $\epsilon_{\rm eff}$. Attempts to use a similar bulk-parameter model for metasurfaces have been less successful. Detailed discussions of this point are given in \cite{hk2} and \cite{awpl}, where it is shown that the surface susceptibilities of a metafilm are the properties that uniquely characterize a metafilm, and as such serve as its most appropriate descriptive parameters. As a result, scattering by a metafilm is best characterized by generalized sheet-transition conditions (GSTCs) \cite{kmh}, in contrast to the effective-medium description used for a metamaterial.  The coefficients appearing in the GSTCs for any given metafilm are all that are required to model its macroscopic interaction with an electromagnetic field.  The GSTCs allow this surface distribution of scatterers to be replaced with a boundary
condition that is applied across an infinitely thin equivalent surface (hence the name metafilm), as indicated in Fig. {\ref{fig2}. The size, shape and spacing of the scatterers are incorporated into this boundary condition through the polarizability densities of the
scatterers on the interface. It was shown in \cite{kmh} that the GSTCs relating the electromagnetic fields on both sides of the metafilm shown in Fig.~\ref{fig1} and Fig.~\ref{fig2} are (under certain conditions, to be discussed below):
\begin{equation}
\textstyle{
\begin{array}{rcl}
\left. {\bf a}_{y}\times\mathbf{E}\right|_{y=0^-}^{0^+} &=&
-j\omega\mu\dyadictall{\boldsymbol{\chi}}_{MS} \cdot \left.\mathbf{H}_{t, {\rm av}}\right|_{y=0}\\
& & - {\bf a}_{y}\times \nabla_{t} \left[ \chi_{ES}^{yy} E_{y, {\rm av}}\right]_{y=0} \\
\left.{\bf a}_{y} \times \mathbf{H}\right|_{y=0^-}^{0^+} &=&
j\omega\epsilon\dyadictall{\boldsymbol{\chi}}_{ES} \cdot \left.\mathbf{E}_{t, {\rm av}}\right|_{y=0}\\
& & - {\bf a}_{y}\times\nabla_{t}\left[
\chi_{MS}^{yy} H_{y, {\rm av}}\right]_{y=0}
\end{array}} \,\,\, ,
 \label{bce}
\end{equation}
where a time dependence $e^{j \omega t}$ has been assumed. The left sides of these expressions represent the jump (or difference) in the tangential components of the fields on the two sides of the metafilm (at $y=0$), and the subscript ``av'' represents the average of the field on either side of the metafilm, i.~e.:
\begin{equation}
\mathbf{E}_{{\rm av}} = \frac{1}{2} \left[ \left. \mathbf{E} \right|_{y=0^+} + \left. \mathbf{E} \right|_{y=0^-} \right] ,
\label{eavv}
\end{equation}
and similarly for the $H$-field. The subscript $t$ refers to components transverse to $y$, and ${\bf a}_y$ denotes the unit vector in the $y$-direction.  The parameters $\dyadictall{\boldsymbol{\chi}}_{ES}$ and $\dyadictall{\boldsymbol{\chi}}_{MS}$ are the dyadic surface electric and magnetic susceptibilities, which have units of meters and are related to the electric and magnetic polarizability densities of the scatterers per unit area. These dyadics vanish when the scatterers are absent, in which case the above boundary conditions reduce to the ordinary condition of continuity of the tangential components of $\mathbf{E}$ and $\mathbf{H}$. The specific type of metafilm analyzed in \cite{kmh} considered only the case where the scatterers and lattice have sufficient symmetry such that the surface susceptibility dyadics are diagonal:
\begin{equation}
\begin{array}{c}
\dyadictall{\boldsymbol{\chi}}_{ES} = \chi_{ES}^{xx} {\bf a}_x {\bf a}_x +
\chi_{ES}^{yy} {\bf a}_y {\bf a}_y + \chi_{ES}^{zz} {\bf a}_z
{\bf a}_z \\
\dyadictall{\boldsymbol{\chi}}_{MS} = \chi_{MS}^{xx} {\bf a}_x {\bf a}_x +
\chi_{MS}^{yy} {\bf a}_y {\bf a}_y + \chi_{MS}^{zz} {\bf a}_z
{\bf a}_z
\end{array}\,\,\, .
\label{chiaa1}
\end{equation}
While this assumption is appropriate for a wide range of metafilms, more general GSTCs for the case of non-symmetric, bi-isotropic, and bi-anisotropic surface susceptibility dyadics are possible.

The surface susceptibility dyadics that appear in the GSTCs are uniquely defined (unlike the thickness and $\mu_{\rm eff}$ and the parameters $\epsilon_{\rm eff}$ that appear when a bulk effective parameter model of a metafilm is attempted). Furthermore, the fields appearing in the GSTCs are ``macroscopic'' fields, in the sense that they exhibit no variations on a length scale comparable to scatterer dimensions or spacing, but only on larger scales such as the wavelength in the surrounding medium.

\begin{figure}
\centering
\scalebox{0.35} {\includegraphics*{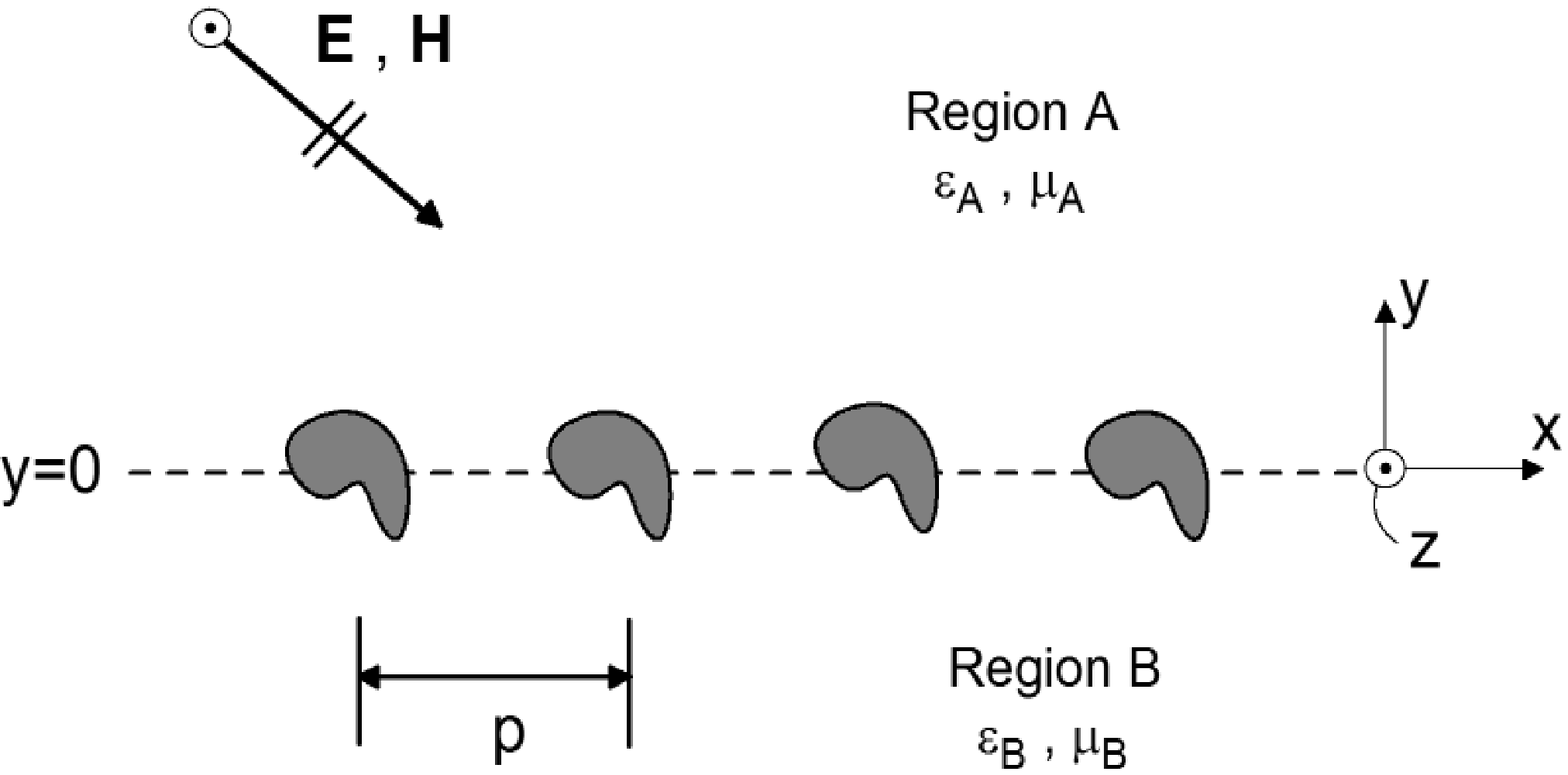}}
\begin{center}
\footnotesize(a)
\end{center}
\centering
\scalebox{0.35} {\includegraphics*{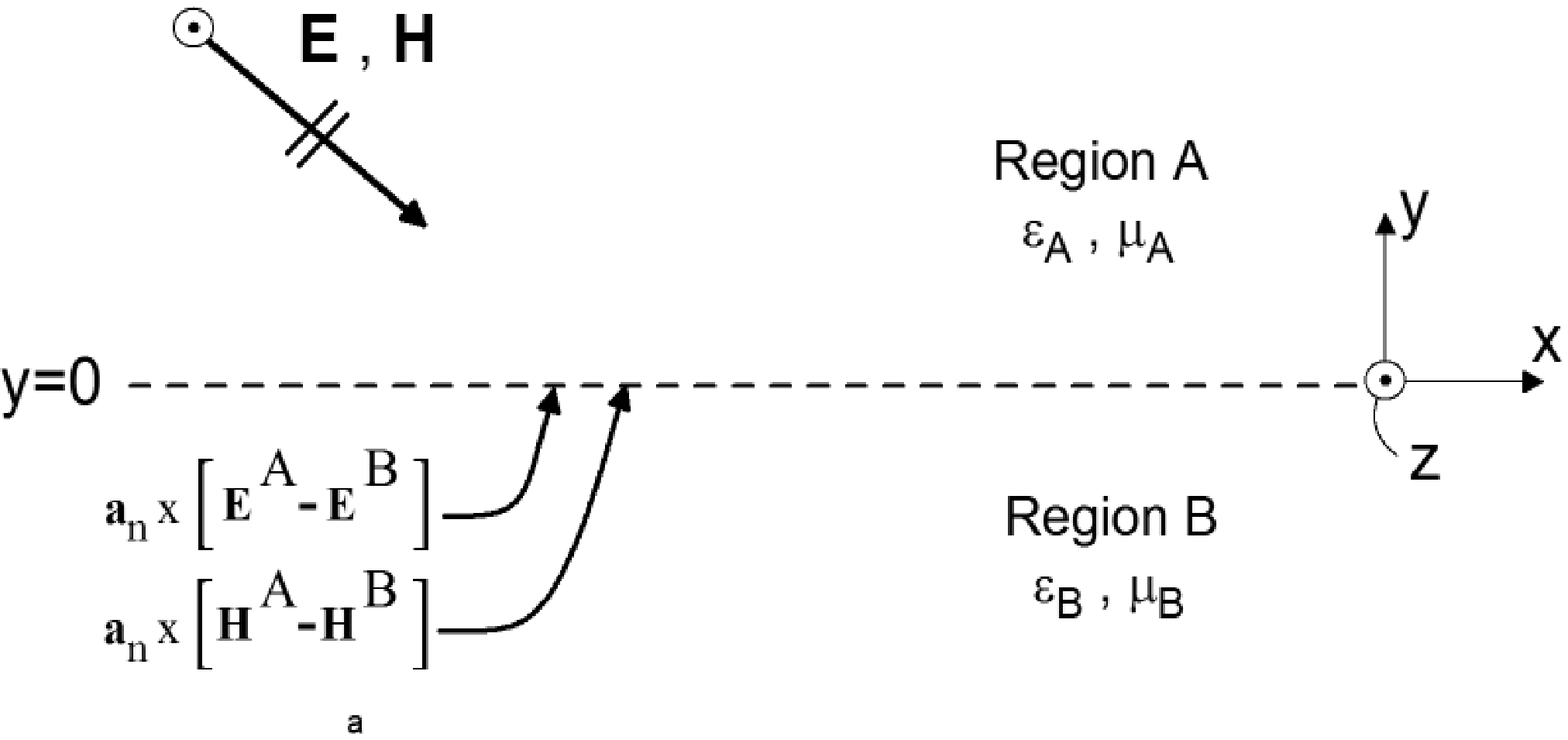}}
\begin{center}
\footnotesize(b)
\end{center}
\caption{(a) Metafilm;  (b) reference plane at which the GSTCs are applied.} \label{fig2}
\end{figure}
\normalsize

Note that in this paper, we refer to the parameters in (\ref{bce}) as ``surface susceptibilities'' (as discussed in \cite{hk3}, \cite{surfacewave}-\cite{awpl}) and use neither the term ``effective surface polarizability densities'' nor the notations $\dyadictall{\boldsymbol{\alpha}}_{ES}$ and $\dyadictall{\boldsymbol{\alpha}}_{MS}$ for them, as was done in \cite{kmh}. This change in terminology was made in order to be less cumbersome and to be consistent with other work \cite{ss1}-\cite{ss3}. When comparing (\ref{bce}) with the GSTCs given in \cite{kmh}, it should be noted that ${\boldsymbol{\chi}}_{MS}$ corresponds to $-\dyadictall{\boldsymbol{\alpha}}_{MS}$, the minus sign originating from the definition of magnetic polarizability used in \cite{kmh}. We should also emphasize that the GSTCs of (1) are appropriate only for metafilms. Metasurfaces with other structures will require a different form of the GSTCs (see \cite{hk3} and \cite{wirehk}).

The GSTCs derived in \cite{kmh} and given in (\ref{bce}) are limited in three ways. First of all, the derivation in \cite{kmh} assumed that the scatterers were in infinite free space, while in the analysis to be carried out in this paper, the scatterers can be embedded in the interface between two different magneto-dielectric media. Second, the derivation in \cite{kmh} assumes only dipole interactions between the scatterers. In doing so, Clausius-Mossotti type models were derived for the surface susceptibilities, assuming that the scatterers are not ``too'' closely spaced.  That assumption will break down if the scatterers become tightly packed.  Finally, the derivation in \cite{kmh} assumes that only diagonal terms appear in the surface susceptibility dyadics as in (\ref{chiaa1}).
For arbitrarily shaped and/or coated scatterers, we could expect off-diagonal terms to appear in these dyadics.  In fact, without giving a formal proof, the authors of \cite{calzo} conjectured that off-diagonal terms should be present in general.  In fact, it has been shown that off-diagonal terms are present in the GSTCs derived for an arbitrarily-shaped coated wire-grating \cite{wirehk}, a similar though related structure.

In this paper, we present a systematic approach based on the technique of multiple-scale homogenization in order to fully characterize the field interaction at the surface of a metafilm shown in Fig. \ref{fig1}.  By this derivation, we will overcome the three limitations of the work in \cite{kmh} noted above. This method will be used to derive GSTCs: equivalent (or ``averaged'') boundary conditions for the metafilm. Due to the geometry of the metafilm, the fields at the interface have both a behavior localized near the scatterers and a global (or average) behavior. The localized field behavior varies on a length scale of the order of the spacing of the scatterers, while the global field behavior varies on a scale of the order of a wavelength. The local field behavior can be separated from that of the average field (through multiple-scale homogenization \cite{wirehk}, \cite{hr1}-\cite{hr7}), representing the field as a product of two functions, one carrying the fine structure and the other the global behavior. A consequence of our analysis is a set of GSTCs for the average or macroscopic field. Hence, the electromagnetic scattering from a metafilm can be approximated by applying the GSTCs at the interface between the two different homogeneous media on either side of the metafilm, as indicated in Figure~\ref{fig2}. These GSTCs, along with Maxwell's equations, are all that is needed to determine macroscopic scattering, transmission, and reflection from the metafilm. If the scale at which information about the field is needed is significantly larger than the fine scale of the system under study, we can discard the information about microscopic field variation, and use only the macroscopic variation of the field (to which only the equivalent boundary condition will apply). If desired, however, the local field behavior can later be reconstructed from the effective fields and associated boundary conditions. In this paper we will show that the homogenization-based derivation results in GSTCs of the same form as those obtained from the dipole interaction model \cite{kmh}, but are not limited to sparsely
spaced scatterers and contain off-diagonal terms for the surface susceptibilities in the case of arbitrarily shaped scatterers.
We note that multiple-scale homogenization has recently been used to analyze some aspects of the electromagnetic problem for thin periodic arrays and layers \cite{delourme1}, \cite{delourme2}, but these authors have not obtained GSTCs, which is the goal of the present paper.

The paper is organized as follows: After the introduction, Section~\ref{s2} lays out the framework of the homogenization technique; we formulate the problem and present the asymptotic expansion of the solution and the boundary conditions that must be satisfied for each term of the expansion. In Section~\ref{s3} the lowest-order terms of the asymptotic expansion are obtained, and in Section~\ref{s4} we solve for the first-order terms in the expansion, from which we derive the GSTCs for the metafilm. Section IV compares results of this paper to those obtained from a Clausius-Mossotti type model. Section~\ref{s5} summarizes the results obtained in the paper, while some details of the derivations are presented in the appendices.

\section{Formulation and Asymptotic Expansions}
\label{s2}
The derivation of the GSTCs is largely analogous to the analysis used in \cite{wirehk}, \cite{hr5}-\cite{hr7}, and we will omit some details when they can be found in these earlier works. This section is divided into several subsections, each covering different aspects of the derivation. The first subsection involves expanding the fields in powers of $k_{0}p$ [where $p$ is the period of the array, $k_0 = \omega \sqrt{\mu_0 \epsilon_0}$ is the free-space wavenumber and $\omega$ is the angular frequency corresponding to an assumed $\exp(j\omega t$) time dependence] and determining boundary conditions for the various field components. Solution of these boundary-value problems will eventually lead to the GSTCs for the effective fields.

\subsection{Asymptotic Expansion of Maxwell's Equations}

Assume that an electromagnetic field is incident onto the array of scatterers as shown in Figs.~\ref{fig1} and \ref{fig2}. For generality, we have assumed that the two media on either side of the metafilm are homogeneous and have different dielectric and magnetic constitutive parameters. In this analysis, we also assume that the scatterers are perfect electric conductors (PEC). However, if we assume that the scatterers are composed of more general materials (i.~e., magneto-dielectric scatterers with either large or small material contrasts), the GSTCs will have the same form and differ only in the specific values of the electric and magnetic surface susceptibilities of the metafilm. In fact, it can be shown that the surface susceptibilities for more general scatterers can exhibit bi-anisotropic properties. By assuming PEC scatterers, we can more easily lay out the essential features of the analysis without
the additional encumbrances that could obscure its understanding.

Since the period $p$ of the array is assumed to be small, there are two spatial length scales, one (the free-space wavelength $\lambda_0$) corresponding to the source or incident wave, and the other ($p$) corresponding to the microstructure of the periodic array of scatterers. The fields will exhibit a multiple-scale type variation that is associated with the microscopic and macroscopic structures of the problem. As in \cite{wirehk}, \cite{hr5}-\cite{hr7}, Maxwell's equations are written as:
\begin{equation}
    \begin{array}{c}
        \nabla{\times}{\bf{E}}^{({\rm A, B})\,T} =
          -j\omega {\bf{B}}^{({\rm A, B})\,T} \,\,\, {\rm :} \,\,\,
         \nabla{\times}{\bf{H}}^{({\rm A, B})\,T}  =
           j\omega {\bf{D}}^{({\rm A, B})\,T} \\
    \end{array}
\label{aa1}
\end{equation}
where
\begin{equation}
\begin{array}{c}
    {\bf{D}}^{({\rm A, B})\,T} = \epsilon_0\epsilon_{r} {\bf{E}}^{({\rm A, B})\,T}\,\,\, {\rm :} \,\,\,
    {\bf{B}}^{({\rm A, B})\,T} = \mu_0\mu_{r} {\bf{H}}^{({\rm A, B})\,T} \,\, ,\\
    \end{array}
\label{aa1a}
\end{equation}
where $T$ indicates the total fields (that contain both the localized and global behaviors), $\mu_{r}$ is the relative permeability and $\epsilon_{r}$ is the relative permittivity at a given observation point. The superscripts A and B denote the regions above and below the plane of the metafilm, respectively.

Continuing as in \cite{hr5}-\cite{hr7}, a multiple-scale representation for the fields in both regions is used:
\begin{equation}
          {\bf{E}}^T({\bf{r}},\mbox{\boldmath$\xi$})=
          {\bf{E}}^T(\frac{{\bf{\hat{r}}}}{ k_0},\bfxi) \,\,\, ,
\label{eh1}
\end{equation}
and similarly for the other fields. Here
\begin{equation}
{\bf{r}}=x{\bf{a}}_x+y{\bf{a}}_y+z{\bf{a}}_z \label{er}
\end{equation}
is the {\it slow} spatial variable, ${\bf{\hat r}}$ is a dimensionless {\it slow} variable given by \cite{hr6}
\begin{equation}
{\bf{\hat r}}=k_0{\bf{r}} \,\,\, , \label{er22}
\end{equation}
and ${\bfxi}$ is a scaled dimensionless variable referred to as the {\it fast} variable and defined as
\begin{equation}
{\bfxi}=\frac{\bf{r}}{p} = \mathbf{a}_x \frac{x}{p} + \mathbf{a}_y \frac{y}{p}+ \mathbf{a}_z \frac{z}{p} = \mathbf{a}_x \xi_x + \mathbf{a}_y \xi_y + \mathbf{a}_z \xi_z \,\,\, , \label{he1}
\end{equation}
where $p$ is the period of the scatterers composing the metafilm, which is assumed to be small compared to all other macroscopic lengths in the problem. The slow variable $\hat{\bf r}$ changes significantly over distances on the order of a wavelength, while the fast variable shows changes over much smaller distances comparable to $p$.

Microscopic variations of the fields in regions A and B with
${\bfxi}$ should be expected close to the array, but once away
from the array this behavior should die out. This suggests a
boundary-layer field representation for the localized terms. The
total fields can thus be expressed in a form making this boundary-layer
effect explicit, as follows:
\begin{equation}
 \mathbf{E}^{T}  = \mathbf{E}(\hat{{\bf{r}}}) + \mathbf{e}(\hat{{\bf{r}}},{\bfxi})\,\,\, ,
\label{blea}
\end{equation}
and similarly for $\mathbf{H}^{T}$. If necessary, we will add a superscript $^{\rmA}$ or $^{\rmB}$ to a field to emphasize that it is to be evaluated in $y>0$ or $y<0$ respectively. The fields  ${\bf{E}_{\rm}}$ and ${\bf{H}_{\rm}}$ are ``non-boundary-layer'' fields, to be referred to henceforth as the effective fields.  The fields $\bf{e}$ and
${\bf{h}}$ are the boundary-layer terms; due to the periodic nature of the array of scatterers, these fields are assumed to be periodic in $\xi_x$ and $\xi_z$ with period 1, but to decay exponentially in $\xi_y$:
\begin{equation}
    \mathbf{e} \,\,\, {\rm and} \,\,\, \mathbf{h}
 = O(e^{-({\rm const})|\xi_y|})
     \,\,\, {\rm as} \,\,\, |\xi_y| \rightarrow \infty \,\,\, .\\
     \label{ehcy}
\end{equation}
Note that the boundary-layer terms are functions of both the {\it fast} and {\it slow} variables. Following similar arguments as in \cite{hr6}, the boundary-layer fields are seen to be functions of only five variables: the slow variables $(\hat{x},\hat{z})$ at the interface that we will represent succinctly by the tangential position vector $\hat{\bf r}_o = {\bf a}_x \hat{x} + {\bf a}_z \hat{z} \equiv k_0 {\bf r}_o$, and $\bfxi$:
\begin{equation}
{\bf{e}}(\hat{\bf r}_o,\bfxi) \label{eehh}\,\,\, .
\end{equation}

To perform the multiple-scale analysis, the del operator must be expressed in terms of the scaled variables and can be represented as \cite{hr6}
\begin{equation}
\nabla \rightarrow k_{0}\nabla_{\hat r} +
\frac{1}{p}\nabla_{\xi}\,\,\, , \label{del}
\end{equation}
where
\begin{equation}
\nabla_{\hat r}={\bf{a}}_{x}\frac{\partial}{\partial \hat{x}} +
           {\bf{a}}_{y}\frac{\partial}{\partial \hat{y}} +
           {\bf{a}}_{z}\frac{\partial}{\partial \hat{z}}
\end{equation}
and
\begin{equation}
\nabla_{\xi}={\bf{a}}_{x}\frac{\partial}{\partial \xi_x} +
           {\bf{a}}_{y}\frac{\partial}{\partial \xi_y}+{\bf{a}}_{z}\frac{\partial}{\partial \xi_z} \,\,\, .
\end{equation}
With the del operator defined in this manner, Maxwell's equations become
\begin{equation}
\begin{array}{rcl}
\nabla_{\hat r}\times{\bf{E}}+\nabla_{\hat r}\times{\bf{e}}+
\frac{1}{\nu}\nabla_{\xi}\times{\bf{e}}&=& -j c \left(
{\bf{B}}+{\bf{b}}\right) \\ \nabla_{\hat
r}\times{\bf{H}}+\nabla_{\hat r}\times{\bf{h}}+
\frac{1}{\nu}\nabla_{\xi}\times{\bf{h}}&=& j c \left(
{\bf{D}}+{\bf{d}}\right) \,\,\, , \\
\end{array}
\label{maxe1}
\end{equation}
where $\nu$ is a small dimensionless parameter defined by
\[
\nu=k_0\,p \,\,\, ,
\]
and $c$ is the speed of light {\em in vacuo}.

We now turn our attention to the relative permeability ($\mu_r$) and relative permittivity ($\epsilon_r$) of the two media, which are given by
\begin{equation}
\epsilon_r = \left\{ \begin{array}{c}
        \epsilon_A \,\, (y > 0) \\
        \epsilon_B \,\, (y < 0)\end{array}\right\} \,\, : \,\,
\mu_r = \left\{ \begin{array}{c}
        \mu_A \,\, (y > 0) \\
        \mu_B \,\, (y < 0)
                     \end{array}
 \right\} \,\,\, ,\\
\label{em1}
\end{equation}
where $\epsilon_{A, B}$ and $\mu_{A, B}$ are the background relative permittivity and permeability of the upper and lower regions A and B, respectively. Note that as defined here they may be discontinuous across the plane $\xi_y = 0$ (i.~e., $y=0$). The reference plane $y=0$ is the dividing line between the two values of background constitutive parameters. It can be chosen to be any convenient position in the boundary-layer, even above or below the metafilm. For simplicity, we will assume that the $y=0$ plane cuts the scatterers composing the metafilm into two parts, see Fig.~\ref{fig3}. A different reference plane location would cause a change in the eventual GSTC obtained, which in
turn would result in a phase shift of reflection and transmission coefficients determined from it. This point is discussed in more detail in \cite{hr6}, \cite{vain}, and \cite{senior}. With this description of the material properties, the constitutive equations (\ref{aa1a}) become
\begin{equation}
\begin{array}{c}
    {\bf{D}}  =  \epsilon_0\epsilon_{r} {\bf{E}} \quad : \quad
    {\bf{B}}  =  \mu_0\mu_{r} {\bf{H}} \,\,\,\, ,\\
       {\bf{d}}  =  \epsilon_0\epsilon_{r} {\bf{e}} \quad : \quad
    {\bf{b}}  =  \mu_0\mu_{r} {\bf{h}} \,\,\, .\\
    \end{array}
\label{aa13}
\end{equation}

Now, the boundary-layer terms of (\ref{maxe1}) vanish by (\ref{ehcy}) as $|\xi_y| \rightarrow \infty$. Thus, the fields away from the metafilm obey the following macroscopic Maxwell equations:
\begin{equation}
   \begin{array}{rcl}
\nabla_{\hat r}\times{\bf{E}} & = & -j c {\bf{B}} \\
   \nabla_{\hat r}\times{\bf{H}} & = & j c {\bf{D}}  \\
   \end{array} \,\,\, .
\label{nbeh}
\end{equation}
But since the effective fields are independent of ${\bfxi}$, equation (\ref{nbeh}) must be true for all $\hat{\bf{r}}$, including up to the plane of the metafilm. Removing the terms of (\ref{nbeh}) from equation (\ref{maxe1}), we
get:
\begin{equation}
   \begin{array}{rcl}
\nabla_{\hat r}\times{\bf{e}} + \frac{1}{\nu} \nabla_{{\xi}}
\times{\bf{e}} & = & -j c {\bf{b}} \\
  \nabla_{\hat r}\times{\bf{h}}
+ \frac{1}{\nu} \nabla_{{\xi}}\times{\bf{h}} & = & j c {\bf{d}}
\,\,\,
. \\
   \end{array}
\label{nbeh2b}
\end{equation}

In our study, we are interested in the case when the period is small compared to a wavelength, which corresponds to $\nu \ll 1$. Thus, it is useful to expand the fields in powers of the small dimensionless parameter $\nu$, i.~e.,
\begin{equation}
   \begin{array}{rcl}
     {\bf{E}} & \sim & {\bf{E}}^0({\bf{r}})+
                \nu{\bf{E}}^{1}({\bf{r}})+O(\nu^{2})  \\
     {\bf{e}} & \sim & {\bf{e}}^0({\bf r}_o,\bfxi)+
                \nu{\bf{e}}^{1}({\bf r}_o,\bfxi)+O(\nu^{2})  \\
    \end{array}
\label{ehsp}
\end{equation}
and similarly for ${\bf H}$, ${\bf h}$ and so forth. The
lowest-order terms (${\bf{E}}^0$, ${\bf{H}}^0$, etc.) include
any incident field which may be present as well as the
zeroth-order scattered field.

Let us now substitute (\ref{ehsp}) into (\ref{nbeh}) and group like powers of $\nu$. We find that each order of effective fields $\mathbf{E}^m$ and $\mathbf{H}^m$ ($m = 0, 1, \ldots$) satisfies the macroscopic Maxwell's equations~(\ref{nbeh}).
Similarly, substituting (\ref{ehsp}) into (\ref{aa13}), and grouping like powers of $\nu$ we obtain the sequence of relations:
\begin{equation}
 \begin{array}{rcl}
\nu^0  & :  &  {\bf{b}}^0  = \mu_0  \mu_r {\bf{h}}^0 \quad {\rm and} \quad
        {\bf{d}}^0  = \epsilon_0 \epsilon_r {\bf{e}}^0\\
  \end{array}
\label{cdo}
\end{equation}
\begin{equation}
 \begin{array}{rcl}
\nu^{1}  & :  &  {\bf{b}}^1  = \mu_0 \mu_r {\bf{h}}^1
\quad {\rm and} \quad  {\bf{d}}^1  = \epsilon_0  \epsilon_r {\bf{e}}^1
  \end{array}
\label{cd1}
\end{equation}
and so on. Furthermore, the different orders of the boundary-layer fields satisfy the following sets of equations; for order $\nu^{-1}$:
\begin{subequations}
\label{do}
 \begin{align}
 \nabla_{\xi}\times{\bf{e}}^0 &= 0 \label{doa} \\
 \nabla_{\xi}\times{\bf{h}}^0 &= 0 \label{dob}
 \end{align}
\end{subequations}
while for order $\nu^m$ ($m = 0, 1, \ldots$):
\begin{subequations}
\label{d1a}
 \begin{align}
 \nabla_{\xi} \times{\bf{e}}^{m+1} &= -jc {\bf{b}}^m -\nabla_{\hat r}\times{\bf{e}}^m \label{d1aa} \\
 \nabla_{\xi} \times{\bf{h}}^{m+1} &= jc {\bf{d}}^m -\nabla_{\hat r}\times{\bf{h}}^m \label{d1ab}
 \end{align}
\end{subequations}
We can understand (\ref{d1a}) to hold also for $m=-1$ if we put $\mathbf{e}^{-1} = 0$ and $\mathbf{h}^{-1} = 0$. By taking the fast divergence $\nabla_{\xi}\cdot$ of (\ref{d1a}) and using some standard vector identities, we have
\begin{equation}
\begin{array}{c}
\nabla_{\xi}\cdot {\bf{b}}^{m+1} =  -\nabla_{\hat{r}} \cdot {\bf b}^m
\quad {\rm and} \quad
 \nabla_{\xi}\cdot {\bf{d}}^{m+1} = -\nabla_{\hat{r}} \cdot {\bf d}^m \,\,\,  \\
\end{array}
\label{he8}
\end{equation}
and specifically
\begin{equation}
  \begin{array}{c}
 \nabla_{\xi}\cdot {\bf{b}}^0  =   0 \quad {\rm and} \quad
 \nabla_{\xi}\cdot {\bf{d}}^0  =  0 \,\,\, . \\
  \end{array}
\label{2ds}
\end{equation}
which, along with~(\ref{d1a}), serve to complete the determination of the higher-order boundary-layer fields. Equations~(\ref{do}) and (\ref{2ds}) show that ${\bf{e}}^0$ and ${\bf{h}}^0$ are static fields that are periodic in $\xi_x$ and $\xi_z$, and decay exponentially as $|\xi_y| \rightarrow \infty$.

From this multiple-scale representation of the fields it is seen
that the effective fields are governed by the macroscopic Maxwell's
equations~(\ref{nbeh}), as expected. On the other hand, the
boundary-layer fields are governed by the static field equations
(\ref{do}) and (\ref{2ds}) at zeroth order, and by (\ref{d1a}) and
(\ref{he8}) at first order.

In order to complete the mathematical definition of the problem, boundary conditions must be specified. When this is done, the effective fields on the metafilm reference plane can be related to the boundary-layer fields at the metafilm interface. In
Section~\ref{s4}, it will be shown that to first order, the
boundary conditions for the effective fields depend only on the
zeroth-order boundary-layer fields. Once the zeroth-order
boundary-layer fields are determined [governed by equations
(\ref{do}) and (\ref{2ds})], the desired first-order boundary
conditions for the effective fields can be obtained.

\subsection{Boundary Conditions at the Interface and on the Scatterers}

The boundary conditions for the fields on the metafilm will now be applied. Before this is done, we will first define the surfaces and boundaries that will be needed in the analysis. In what follows, various integrations will be performed over portions of the periodic unit-cell shown in Figs.~\ref{fig3} and \ref{fig4}.
\begin{figure}[t!]
\centering
\scalebox{0.35}{\includegraphics*{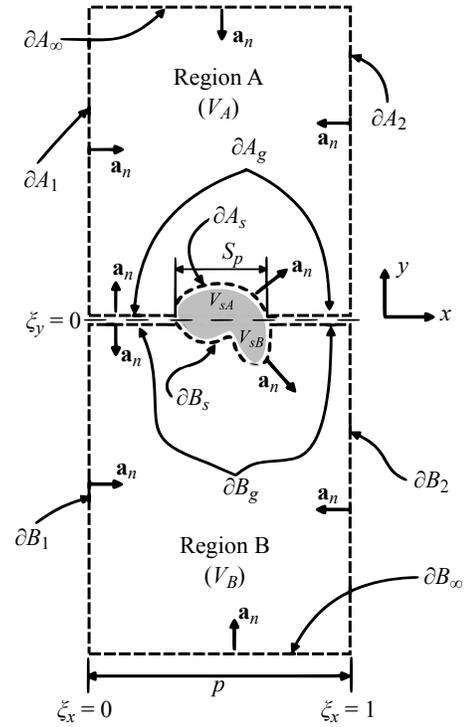}}
\caption{Cross-section of the period cell at a plane $\xi_z =$ constant.} \label{fig3}
\end{figure}
\begin{figure}[t!]
\centering
\scalebox{0.52}{\includegraphics*{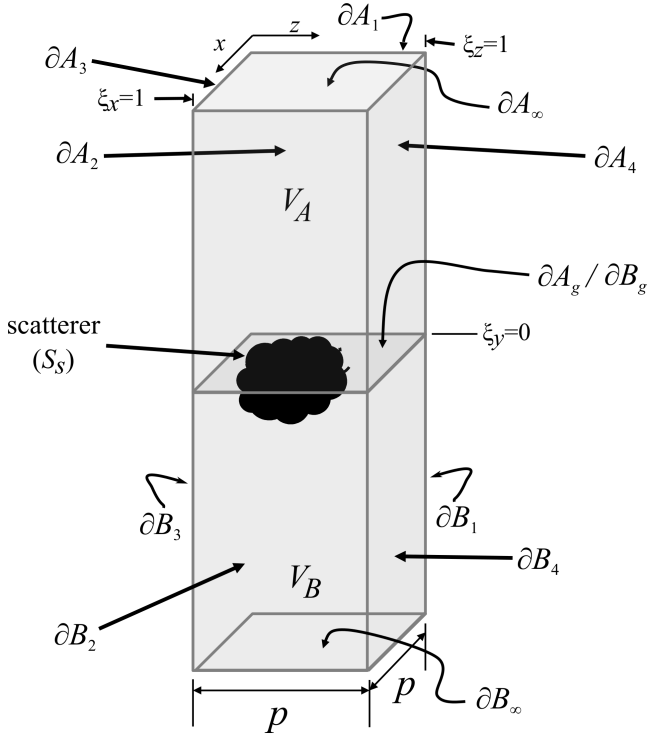}}
\caption{Three-dimensional view of the period cell.} \label{fig4}
\end{figure}
Regions $V_A$ and $V_B$ are the portions of the unit cell lying outside the scatterer surface $S_s$, in $\xi_y >0$ or $\xi_y <0$ respectively. The interior of the scatterer is denoted by $V_s$, which is divided into the portions $V_{sA}$ and $V_{sB}$ lying above and below $\xi_y = 0$, respectively. The entire volume of the period cell exterior to the scatterer will be denoted as $V = V_A \cup V_B$. The boundaries of these regions are denoted by $\partial A$ and $\partial B$, respectively, with the unit normal vector ${\bf{a}}_n$ always taken ``into'' region $V_A$ or $V_B$; in particular,
\begin{equation}
{\bf{a}}_n|_{\partial A_g}= -{\bf{a}}_n|_{\partial B_g} = \mathbf{a}_y
\end{equation}
in the gap portion of the plane $\xi_y = 0$, denoted by $\partial A_{g}$ or $\partial B_{g}$. The other portions of $\partial A$ and $\partial B$ are: the portions $\partial A_s$ and $\partial B_s$ of the boundary $S_s$ of the PEC scatterer that lie in $V_A$ or $V_B$ respectively (Fig.~\ref{fig3}); hence $S_s = \partial A_s \cup \partial B_s$. Our convention for ${\bf{a}}_n$ means that it is always directed outward from $S_s$. The remaining portions of $\partial A$ and $\partial B$ are the sidewalls of the period cell:
\begin{eqnarray}
S_1 = \partial A_1 \cup \partial B_1 , && S_2 = \partial A_2 \cup \partial B_2 , \nonumber \\
S_3 = \partial A_3 \cup \partial B_3 , && S_4 = \partial A_4 \cup \partial B_4 ,
 \label{s1s4}
\end{eqnarray}
where
\begin{eqnarray}
 \partial A_1 & : & \mbox{\rm ($\xi_x = 0$, $\xi_y > 0$ and $0 < \xi_z <1$)} \nonumber \\
 \partial A_2 & : & \mbox{\rm ($\xi_x = 1$, $\xi_y > 0$ and $0 < \xi_z <1$)} \nonumber \\
 \partial A_3 & : & \mbox{\rm ($0 < \xi_x <1$, $\xi_y > 0$ and $\xi_z = 0$)} \nonumber \\
 \partial A_4 & : & \mbox{\rm ($0 < \xi_x <1$, $\xi_y > 0$ and $\xi_z = 1$)} \nonumber \\
 \partial B_1 & : & \mbox{\rm ($\xi_x = 0$, $\xi_y < 0$ and $0 < \xi_z <1$)} \nonumber \\
 \partial B_2 & : & \mbox{\rm ($\xi_x = 1$, $\xi_y < 0$ and $0 < \xi_z <1$)} \nonumber \\
 \partial B_3 & : & \mbox{\rm ($0 < \xi_x <1$, $\xi_y < 0$ and $\xi_z = 0$)} \nonumber \\
 \partial B_4 & : & \mbox{\rm ($0 < \xi_x <1$, $\xi_y < 0$ and $\xi_z = 1$)} \nonumber
\end{eqnarray}

Various boundary conditions hold on the different portions of the boundaries of these regions. The boundary-layer fields $\mathbf{e}^m$ and $\mathbf{h}^m$ must decay exponentially to zero on $\partial A_{\infty}$ (corresponding to the boundary where $\xi_y\rightarrow\infty$), and on $\partial B_{\infty}$ (where $\xi_y\rightarrow-\infty$). They must also be periodic in $\xi_x$ and $\xi_z$.

On the remaining parts of the boundary, let us consider the ${\bf E}$ field first. In the gap $\partial A_{g}$ or $\partial B_{g}$ the total tangential field is continuous, while on the surface of the scatterers the total tangential $E$-field is zero:
\begin{equation}
\left. {\bf{a}}_{n}\times{\bf{E}}^{\rmA,\,T}\right|_{\partial A_g}=-
\left. {\bf{a}}_{n}\times{\bf{E}}^{\rmB,\,T}\right|_{\partial B_g} \,\,
, \label{etgap}
\end{equation}
and
\begin{equation}
\left. {\bf{a}}_{n}\times{\bf{E}}^{\rmA,\,T}\right|_{\partial A_s}=
\left. {\bf{a}}_{n}\times{\bf{E}}^{\rmB,\,T}\right|_{\partial
B_s}\equiv 0 \,\,\, .\label{etcon}
\end{equation}
We can evaluate the effective fields appearing in the expression for the fields on the scatterers by extrapolation
relative to the reference plane $y=0$ (Fig.~\ref{fig3}) using a Taylor series in $y$. Any function of the slow variables only can thus be expanded in the boundary layer as:
\begin{equation}
f({\bf{r}}) = f(x,0,z) + \nu \xi_y \left. \frac{{\partial}f(x,y,z)}{{\partial}{\hat y}} \right|_{y=0} +O(\nu^{2})\,\,\, ,
\label{taylor}
\end{equation}
where $\hat{y}=k_0 y=\nu \xi_y$ was used.  Using expansions (\ref{taylor}) and (\ref{ehsp}), equation (\ref{etcon}) can be expanded up to terms of order $\nu$ to give the boundary conditions for the tangential $E$-field on $\partial A_s$ as
\begin{equation}
\nu^0  :
\left.{\bf{a}}_{n}\times{\bf{e}}^{\rmA 0}\right|_{{\partial}A_{s}}=
-{\bf{a}}_{n}\times{\bf{E}}^{\rmA 0}({\bf r}_o)
\label{dobc}
\end{equation}
\begin{equation}
\nu^{1}:
\left.{\bf{a}}_{n}\times{\bf{e}}^{\rmA 1}\right|_{{\partial}A_{s}}=
-\xi_y{\bf{a}}_{n}\times \left[\frac{\partial }{{\partial}{\hat y}}
{\bf{E}}^{\rmA 0}\right]_{y=0}
-{\bf{a}}_{n}\times {\bf{E}}^{\rmA 1}({\bf
r}_o).
\label{d1bc}
\end{equation}
and so on. Likewise, on $\partial B_s$ we have
\begin{equation}
 \nu^0  :
\left.{\bf{a}}_{n}\times{\bf{e}}^{\rmB 0}\right|_{{\partial}B_{s}}=
-{\bf{a}}_{n}\times{\bf{E}}^{\rmB 0}({\bf r}_o)
\label{dobc2}
\end{equation}
\begin{equation}
\nu^{1} :
\left.{\bf{a}}_{n}\times{\bf{e}}^{\rmB 1}\right|_{{\partial}B_{s}}=
-\xi_y{\bf{a}}_{n}\times \left[\frac{\partial}{{\partial}{\hat y}}
{\bf{E}}^{\rmB 0}\right]_{y=0}-{\bf{a}}_{n}\times {\bf{E}}^{\rmB 1}({\bf
r}_o)  \,\,\, , \\
\label{d1bc2}
\end{equation}
Using (\ref{ehsp}) and (\ref{etgap}), in the gap [denoted by ($\partial A_g/\partial B_g$)] we have
\begin{equation}
{\bf{a}}_{y}\times\left[{\bf{e}}^{\rmA m}-{\bf{e}}^{\rmB m}\right]_{{\partial}A_{g}/\partial B_g}= -{\bf{a}}_{y}\times\left[{\bf{E}}^{\rmA m}({\bf
r}_o)-{\bf{E}}^{\rmB m}({\bf r}_o)\right] \\
\label{dobc3}
\end{equation}
where $m=0, 1, \ldots$ denotes the order of the field in expansion (\ref{ehsp}). The continuity of the total tangential $H$ in the gap gives
\begin{equation}
{\bf{a}}_{y}\times\left[{\bf{h}}^{\rmA m}-{\bf{h}}^{\rmB m}\right]_{{{{\partial}A_{g}/\partial B_g}}}= -{\bf{a}}_{y}\times\left[{\bf{H}}^{\rmA m}({\bf
r}_o)-{\bf{H}}^{\rmB m}({\bf r}_o)\right].\\
\label{ht0bc}
\end{equation}

Static problems require boundary conditions on both tangential and normal field components to ensure uniqueness (except when unknown surface charges and current are involved). Thus, in the gaps we must also impose that the normal component of the total $D$-field is continuous:
\begin{equation}
\left. {\bf{a}}_{n}\cdot{\bf{D}}^{\rmA,\,T}\right|_{\partial A_g}=-
\left. {\bf{a}}_{n}\cdot{\bf{D}}^{\rmB,\,T}\right|_{\partial B_g} \,\,
, \label{engap} \end{equation}
from which we get
\begin{equation}
\left.{\bf{a}}_{y}\cdot\left[{\bf{d}}^{\rmA m}-{\bf{d}}^{\rmB m}\right]\right|_{{\partial}A_{g}/\partial B_{g}}=
-{\bf{a}}_{y}\cdot\left[{\bf{D}}^{\rmA m}({\bf r}_o)-{\bf{D}}^{\rmB m}({\bf r}_o)\right] \, .
\label{dn0bc}
\end{equation}
Likewise, the normal component of the total $B$-field on the scatterers is zero, and across the gaps between the PEC scatterers it is continuous; in the gap we have
\begin{equation}
\left. {\bf{a}}_{y}\cdot{\bf{B}}^{\rmA,\,T}\right|_{\partial A_g}=-
\left. {\bf{a}}_{y}\cdot{\bf{B}}^{\rmB,\,T}\right|_{\partial B_g} \,\,
, \label{bngap} \end{equation}
while on the scatterers
\begin{equation}
\left. {\bf{a}}_{n}\cdot{\bf{B}}^{\rmA,\,T}\right|_{\partial A_s}=
\left. {\bf{a}}_{n}\cdot{\bf{B}}^{\rmB,\,T}\right|_{\partial B_s}\equiv
0 \,\,\, .
\label{bncon}
\end{equation}
On $\partial A_s$, this gives
\begin{equation}
\nu^0 :
\left.{\bf{a}}_{n}\cdot{\bf{b}}^{\rmA 0}\right|_{{\partial}A_{s}}=
-{\bf{a}}_{n}\cdot{\bf{B}}^{\rmA 0}({\bf r}_o) \\
\label{dobc33}
\end{equation}
\begin{equation}
\nu^{1} :
\left.{\bf{a}}_{n}\cdot{\bf{b}}^{\rmA 1}\right|_{{\partial}A_{s}}=
-\xi_y{\bf{a}}_{n}\cdot \left[\frac{\partial}{{\partial}{\hat y}}
{\bf{B}}^{\rmA 0}\right]_{y=0}-{\bf{a}}_{n}\cdot {\bf{B}}^{\rmA 1}({\bf
r}_o)  \,\,\, . \\
\label{d1bcbb}
\end{equation}
and on $\partial B_s$ we have
\begin{equation}
\nu^0 :
\left.{\bf{a}}_{n}\cdot{\bf{b}}^{\rmB 0}\right|_{{\partial}B_{s}}=
-{\bf{a}}_{n}\cdot{\bf{B}}^{\rmB 0}({\bf r}_o) \\
\label{dobc23}
\end{equation}
\begin{equation}
 \nu^{1} :
 \left.{\bf{a}}_{n}\cdot{\bf{b}}^{\rmB 1}\right|_{{\partial}B_{s}}=
-\xi_y{\bf{a}}_{n}\cdot \left[\frac{\partial}{{\partial}{\hat y}}
{\bf{B}}^{\rmB 0}\right]_{y=0}-{\bf{a}}_{n}\cdot {\bf{B}}^{\rmB 1}({\bf
r}_o)\  \, , \\
\label{d1bc23}
\end{equation}
while in the gap we have
\begin{equation}
\left.{\bf{a}}_{y}\cdot\left[{\bf{b}}^{\rmA m}-{\bf{b}}^{\rmB m}\right]\right|_{{\partial}A_{g}/\partial
B_{g}}= -{\bf{a}}_{y}\cdot\left[{\bf{B}}^{\rmA m}({\bf
r}_o)-{\bf{B}}^{\rmB m}({\bf r}_o)\right] \, .\\
\label{dobc3b}
\end{equation}

\subsection{Continuity Conditions on the Zeroth-Order Effective Fields at the Reference Surface}
\label{s22}

The solvability constraints obtained in Appendix~\ref{apbl} can be used to obtain continuity conditions on the macroscopic fields. Using (\ref{solvem0}) and (\ref{doa}) we have the first of the desired boundary conditions for the zeroth-order electric field:
\begin{equation}
{\bf{a}}_y\times\left[{\bf{E}}^{\rmA 0}({\bf r}_o) -{\bf{E}}^{\rmB 0}({\bf r}_o)\right] =0 \,\,\, .
\label{bceo1}
\end{equation}
In a similar way, from (\ref{solvhm0}) and (\ref{dob}) we get:
\begin{equation}
{\bf{a}}_y\times\left[{\bf{H}}^{\rmA 0}({\bf r}_o) -{\bf{H}}^{\rmB 0}({\bf r}_o)\right] =0 \,\,\, . \\
\label{bcho1}
\end{equation}
From the solvability conditions (\ref{solvdm0}) and (\ref{solvbm0}) together with (\ref{2ds}), we obtain the continuity of the normal components of ${\bf{B}}^0$ and ${\bf{D}}^0$ at $y=0$:
\begin{equation}
{\bf{a}}_{y}\cdot\left[{\bf{D}}^{\rmA 0}({\bf r}_o)-{\bf{D}}^{\rmB 0}({\bf r}_o)\right]=0\,\,\,.
\label{sur4e}
\end{equation}
\begin{equation}
{\bf{a}}_{y}\cdot\left[{\bf{B}}^{\rmA 0}({\bf r}_o)-{\bf{B}}^{\rmB 0}({\bf r}_o)\right]=0
\label{sur4b}
\end{equation}
To zeroth order, the tangential components of the effective $E$ and $H$-fields and the normal components of $D$ and $B$ are continuous across the metafilm, just as they are at an ordinary material interface in the absence of surface current and charge densities.

\section{Derivation of the GSTCs}

In this section, we will derive generalized transfer-type boundary conditions for the effective fields at the reference surface $y=0$ as defined in Fig.~\ref{fig2}.
The derivation will be based on some integral identities derived in the appendices. We first explicitly state the governing equations for the zeroth-order boundary layer fields and introduce some normalized boundary-layer fields. Then, we express the effective fields at this surface in terms of surface integrals of these zeroth-order normalized boundary-layer fields. These integrals are finally evaluated so as to obtain the desired GSTCs.

\subsection{Lowest-Order Boundary-Layer Fields}
\label{s3}

With the fields separated into effective and boundary layer terms (that obey the appropriate differential equations and boundary conditions), it is possible to analyze them individually at each order of $\nu$. The zeroth-order boundary-layer fields ${\bf{e}}^0$ and ${\bf{h}}^0$ are of particular importance because integrals of these fields will turn out to be directly related to the surface susceptibilities that characterize the metafilm. They are governed by (\ref{do}) and (\ref{2ds}), together with the relevant boundary conditions, which for convenience we gather together here for the electric field:
\begin{subequations}
\label{peo}
 \begin{align}
 \mbox{\rm for ${\bfxi}\,\in\,\, V$:} \qquad \ \, \ \ \nabla_{\xi}\times{\bf{e}}^0 &=0 \label{peoa} \\
 \mbox{\rm for ${\bfxi}\,\in\,\, V$:} \qquad \nabla_{\xi}\cdot \left( \epsilon_{r} {\bf{e}}^0 \right) &= 0 \label{peob} \\
 {\bf{a}}_{n} \times \left[ {\bf{e}}^{\rmA 0} + {\bf{E}}^{\rmA 0}({\bf r}_o) \right]_{{\partial}A_{s}} &= 0 \label{peoc} \\
 {\bf{a}}_{n} \times \left[ {\bf{e}}^{\rmB 0} + {\bf{E}}^{\rmB 0}({\bf r}_o) \right]_{{\partial}B_{s}} &= 0 \label{peod} \\
 {\bf{a}}_{y} \times \left[{\bf{e}}^{\rmA 0}-{\bf{e}}^{\rmB 0}\right]_{{\partial}A_{g}/\partial B_{g}} &= 0 \label{peoe} \\
 {\bf{a}}_{y} \cdot \left[{\bf{d}}^{\rmA 0}-{\bf{d}}^{\rmB 0}\right]_{{\partial}A_{g}/\partial B_{g}} &= 0 \label{peof}
  \end{align}
\end{subequations}
and for the magnetic field:
\begin{subequations}
\label{pho}
 \begin{align}
 \mbox{\rm for ${\bfxi}\,\in\,\, V$:} \qquad \kern0.5pt \ \ \ \ \nabla_{\xi}\times{\bf{h}}^0 &=0 \label{phoa} \\
 \mbox{\rm for ${\bfxi}\,\in\,\, V$:} \qquad \nabla_{\xi}\cdot \left( \mu_{r} {\bf{h}}^0 \right) &= 0 \label{phob} \\
 {\bf{a}}_{n} \cdot \left[ {\bf{b}}^{\rmA 0} + {\bf{B}}^{\rmA 0}({\bf r}_o) \right]_{{\partial}A_{s}} &= 0 \label{phoc} \\
 {\bf{a}}_{n} \cdot \left[ {\bf{b}}^{\rmB 0} + {\bf{B}}^{\rmB 0}({\bf r}_o) \right]_{{\partial}B_{s}} &= 0 \label{phod} \\
 {\bf{a}}_{y} \times \left[{\bf{h}}^{\rmA 0}-{\bf{h}}^{\rmB 0}\right]_{{\partial}A_{g}/\partial B_{g}} &= 0 \label{phoe} \\
 {\bf{a}}_{y} \cdot \left[{\bf{b}}^{\rmA 0}-{\bf{b}}^{\rmB 0}\right]_{{\partial}A_{g}/\partial B_{g}} &= 0 \label{phof}
  \end{align}
\end{subequations}

It will be useful to express the zeroth-order boundary-layer fields in terms of the effective fields at $y=0$. The boundary conditions for the zeroth-order boundary-layer fields contain only the macroscopic fields at the reference plane $y=0$ as sources (forcing terms), so the boundary-layer fields will be proportional to these forcing terms. Since the zeroth-order effective $E_x$, $E_z$ and $D_y$ are continuous at the interface, we may omit the superscript A or B on these. From equations (\ref{peo}) we can see that the sources for ${\bf{e}}^0$ are ${E}^0_{x}(\mathbf{r}_o)$, ${D}^0_{y}(\mathbf{r}_o)$, and ${E}^0_{z}(\mathbf{r}_o)$, with an analogous statement holding for $\mathbf{h}^0$. By superposition, we see that ${\bf{e}}^0$ and ${\bf{h}}^0$
must therefore have the following form:
\begin{equation}
{\bf{e}}^0 = {E}^0_{x}(\mathbf{r}_o){\sce}_{1}(\bfxi) + \frac{D^0_{y}(\mathbf{r}_o)}{\epsilon_0}{\sce}_{2}(\bfxi) + {E}^0_{z}(\mathbf{r}_o){\sce}_{3}(\bfxi)\,\, ,
\label{dsce}
\end{equation}
\begin{equation}
{\bf{h}}^0 = {H}^0_{x}(\mathbf{r}_o){\sch}_{1}(\bfxi) + \frac{B^0_{y}(\mathbf{r}_o)}{\mu_0} {\sch}_{2}(\bfxi) + {H}^0_{z}(\mathbf{r}_o){\sch}_{3}(\bfxi)\,\, ,
\label{dsch}
\end{equation}
where ${\sce}_{i}$ and ${\sch}_{i}$ are dimensionless functions of the fast variables only, the governing equations needed for whose determination are given in Appendix~\ref{apa}. Hereafter, it will sometimes be convenient to use numerical indices $i$ or $k = 1,2,3$ to denote the coordinates $x$, $y$ or $z$ respectively. Thus, $\mathbf{a}_1 = \mathbf{a}_x$, $\mathbf{a}_2 = {\bf{a}}_y$, and $\mathbf{a}_3 = {\bf{a}}_z$. The subscript $i$ in $\sce_i$ or $\sch_i$ indicates the component of the macroscopic ``source'' field that produces it.

Using the representations (\ref{dsce}) and (\ref{dsch}) of ${\bf{e}}^0$ and ${\bf{h}}^0$, the curl with respect to the slow spatial variable of each of these fields is expressed as
\begin{equation}
  \begin{array}{c}
\nabla_{\hat r}\times{\bf{e}}^{0}  =
          -{\sce}_1\times\nabla_{t,\hat r}{E}^{0}_{x}({\bf r}_o)
          - \frac{1}{\epsilon_0} {\sce}_2\times\nabla_{t,\hat r}{D}^{0}_{y}({\bf r}_o)\\
-{\sce}_3\times\nabla_{t,\hat r}{E}^{0}_{z}({\bf r}_o) \\

\nabla_{\hat r}\times{\bf{h}}^{0}  =
         -{\sch}_1\times\nabla_{t,\hat r}
{H}^{0}_{x}({\bf r}_o)
          - \frac{1}{\mu_0} {\sch}_2\times\nabla_{t,\hat r}
{B}^{0}_{y}({\bf r}_o)\\
-{\sch}_3\times\nabla_{t,\hat r}
{H}^{0}_{z}({\bf r}_o)\,\,\,\, .\\
 \end{array}
\label{curlslow}
\end{equation}
The subscript ``$t$'' corresponds to derivatives with respect to $x$ and $z$ only: ${\bf{e}}$ and ${\bf{h}}$ are independent of $y$, so the curl expressions on the left hand side of (\ref{curlslow}) contain no $y$-derivatives.

\subsection{Solvability Conditions for the First-Order Fields and the GSTCs}
\label{s4}
Thus far, we have obtained boundary conditions only for the zeroth-order effective fields. In this section, the first-order effective fields are investigated, and the essential boundary conditions for them are derived by enforcing solvability conditions on the first-order boundary-layer fields. These results will then be used to obtain the GSTCs.

We start by applying (\ref{solvem1}) and (\ref{d1aa}) to get:
\begin{eqnarray}
\lefteqn{{\bf{a}}_{y}\times\left[{\bf{E}}^{\rmA 1}({\bf r}_o)-{\bf{E}}^{\rmB 1}({\bf r}_o)\right] =} \label{lside2t} \\
&& - {\bf{a}}_y\times \left[ \hat{V}_{sA} \frac{\partial \bf{E}^{\rmA0}}{\partial {\hat y}} + \hat{V}_{sB} \frac{\partial \bf{E}^{\rmB 0}}{\partial {\hat y}} \right]_{y=0} \nonumber \\
&& - jc \int_V \mathbf{b}^0 \, dV - \nabla_{\hat{r}} \times \int_V \mathbf{e}^0 \, dV \nonumber
\end{eqnarray}
where $\hat{V}_{sA}$ and $\hat{V}_{sB}$ are the scaled volumes of the scatterer that are above and below the $\xi_y=0$ (see Appendix~\ref{ap0}). All integrals in this paper are understood to be with respect to the fast variable $\bfxi$. But from the components of Faraday's law transverse to $y$, we have
\begin{equation}
\begin{array}{c}
{\bf a}_y\times\left. \frac{\partial \bf{E}^0}{\partial {\hat y}}\right|_{{\bf r}_o} =
-j\eta_0\mu_r \left[{\bf{a}}_x H_x^0({\bf r}_o)+{\bf{a}}_z H_z^0({\bf r}_o) \right] \\
+ \frac{1}{\epsilon_0 \epsilon_r} {\bf a}_y \times \nabla_{t,\hat r} {{D}}^0_y({\bf{r}}_0) \,\,\, ,\\
\end{array}
\label{fe3}
\end{equation}
where $\eta_0=\sqrt{\mu_0/\epsilon_0}$ is the free space wave impedance, so expression~(\ref{lside2t}) becomes
\begin{eqnarray}
\lefteqn{{\bf{a}}_{y}\times\left[{\bf{E}}^{\rmA 1}({\bf r}_o)-{\bf{E}}^{\rmB 1}({\bf r}_o)\right] =} \label{lside2} \\
&& j\eta_0 \left(\mu_A\,\hat{V}_{sA}+\mu_B\,\hat{V}_{sB}\right) \mathbf{H}_t^0({\bf r}_o) \nonumber \\
&& - \frac{1}{\epsilon_0} \left( \frac{\hat{V}_{sA}}{\epsilon_A} + \frac{\hat{V}_{sB}}{\epsilon_B} \right) {\bf{a}}_y \times\nabla_{t,{\hat{r}}} D_y^0({\bf r}_o) \nonumber \\
&& - jc \int_V \mathbf{b}^0 \, dV - \nabla_{\hat{r}} \times \int_V \mathbf{e}^0 \, dV \nonumber
\end{eqnarray}
Using (\ref{dsce}) and (\ref{dsch}), the two integrals is this expression become
\begin{equation}
\begin{array}{c}
- jc \int_V \mathbf{b}^0 \, dV - \nabla_{\hat{r}} \times \int_V \mathbf{e}^0 \, dV  = \\
  -j\eta_0 \int_{V_{AB}}\mu_r \left[ {H}^0_{x}({\mathbf{r}}_o) {\sch}_{1} + \frac{{B}^0_{y}({\mathbf{r}}_o)}{\mu_0} {\sch}_{2} + {H}^0_{z}({\mathbf{r}}_o) {\sch}_{3} \right] \, dV_{\xi} \\
+\int_{V_{AB}}{\sce}_{1} dV\times\nabla_{t,\hat r}  {E}^0_{x}(\mathbf{r}_o) + \int_{V_{AB}}{\sce}_{2} dV\times\nabla_{t,\hat r}  \frac{D^0_{y}(\mathbf{r}_o)}{\epsilon_0} \\
+\int_{V_{AB}}{\sce}_{3} dV\times\nabla_{t,\hat r}  {E}^0_{z}(\mathbf{r}_o) \,\,\,\, .\\
\end{array}
\label{bccurle1}
\end{equation}
Using procedures similar to those in Appendix~C of \cite{wirehk} it can be shown that $\int\sce_2\, dV$ has only a $y$-component while $\int\sch_{1,3}\, dV$ have no $y$-components. With this and the fact that the $y$-component of Faraday's Law for the zeroth-order effective field:
\begin{equation}
j c B_y^0({\bf{r}}_0) + \mathbf{a}_y \cdot \nabla_{\hat{r}} \times \mathbf{E}_t^0 = 0 ,
\label{fe2}
\end{equation}
the $y$-components of (\ref{lside2}) can be shown to cancel. The jump in the first-order effective $E$-field across the metafilm becomes
\begin{equation}
\begin{array}{c}
{{\bf{a}}_y\times\left[{\bf{E}}^{\rmA 1}({\bf r}_o) -{\bf{E}}^{\rmB 1}({\bf r}_0)\right] =}  \\
 -{\bf{a}}_x\,j\eta_0\,\left[
 \frac{\hat{\chi}_{MS}^{xx}}{p} {H}^0_{x}({\mathbf{r}}_o)
+ \frac{\hat{\chi}_{MS}^{xy}}{p}  \frac{{B}^0_{y}}{\mu_0} ({\mathbf{r}}_o)
+ \frac{\hat{\chi}_{MS}^{xz}}{p} {H}^0_{z}({\mathbf{r}}_o)
 \right] \\
 -{\bf{a}}_z\,j\eta_0\,\left[
 \frac{\hat{\chi}_{MS}^{zx}}{p} {H}^0_{x}({\mathbf{r}}_o)
+ \frac{\hat{\chi}_{MS}^{zy}}{p}  \frac{{B}^0_{y}}{\mu_0} ({\mathbf{r}}_o)
+\frac{ \hat{\chi}_{MS}^{zz}}{p} {H}^0_{z}({\mathbf{r}}_o)
 \right]  \\
 - {\bf{a}}_y\times \nabla_{t,\hat r} \left[  \frac{\hat{\chi}_{ES}^{yx}}{p} {E}^{0}_{x} +\frac{ \hat{\chi}_{ES}^{yy} }{p} \frac{D^0_{y}(\mathbf{r}_o)}{\epsilon_0} + \frac{\hat{\chi}_{ES}^{yz}}{p} {E}^{0}_{z}\right]
\end{array}
\label{bce1effta}
\end{equation}
where the coefficients $\chi_{ES}$ and $\chi_{MS}$  are define in terms of the various integrals in (\ref{bccurle1}) and are given in (\ref{chiEE}) and (\ref{chiMM}), see Appendix~\ref{aplast}. These coefficients have units of meters and are interpreted as effective electric and magnetic surface susceptibilities of the metafilm.

We now turn to the derivation of a jump condition for the first-order tangential $H$-field, and we start by applying (\ref{solvhm1}) and (\ref{d1ab}):
\begin{eqnarray}
\lefteqn{{\bf{a}}_{y}\times\left[{\bf{H}}^{\rmA 1}({\bf r}_o)-{\bf{H}}^{\rmB 1}({\bf r}_o)\right] =} \label{bch2} \\
&& j c \int_{V} {\bf{d}}^{0} \, dV+j c  \oint_{S_s}{\bfxi}\,{\bf{a}}_{n}\cdot{\bf{d}}^0\,dS \nonumber \\
&& - \nabla_{t,\hat r}\times \int_{V} {\bf{h}}^{0} \, dV - \oint_{S_s}{\bfxi}\,{\bf{a}}_{n} \cdot \left(\nabla_{t,\hat{r}}\times{\bf{h}}^{0}\right) \,dS \nonumber \\
&& -{\bf{a}}_{y}\times \left[ \hat{V}_{sA}\, \frac{\partial {\bf{H}}^{\rmA 0}}{{\partial}{\hat y}} + \hat{V}_{sB}\, \frac{\partial {\bf{H}}^{\rmB 0}}{{\partial}{\hat y}}
 \right]_{y=0} \,\,\, . \nonumber
\end{eqnarray}
Using (\ref{pyn0}) for the $y$-components and (\ref{pynx}) for the transverse components, the first and second terms on the right side of (\ref{bch2}) can be rewritten to give:
\begin{eqnarray}
\lefteqn{\int_{V} {\bf{d}}^{0} \, dV +  \oint_{S_s}{\bfxi}\,{\bf{a}}_{n} \cdot{\bf{d}}^0\,dS =} \label{sumne} \\
 && \mathbf{a}_x \int_{S_2} \mathbf{a}_x \cdot \mathbf{d}^0\, dS + \mathbf{a}_z \int_{S_4} \mathbf{a}_z \cdot \mathbf{d}^0\, dS \nonumber
\end{eqnarray}
where the surfaces $S_2$ and $S_4$ are defined in (\ref{s1s4}). Taking the slow divergence of (\ref{mynx}) and using some vector identities, the third and fourth terms on the right side of (\ref{bch2}) can be rewritten, giving:
\begin{equation}
\begin{array}{c}
- \nabla_{t,\hat r}\times \int_{V} {\bf{h}}^{0} \, dV - \oint_{S_s}{\bfxi}\,{\bf{a}}_{n} \cdot \left(\nabla_{\hat{r}}\times{\bf{h}}^{0}\right) \,dS= \\
{\bf a}_x \nabla_{t,\hat{r}} \cdot \left[ {\bf a}_x \times \int_{S_2} {\bf{h}}^{0} \, dS \right]
+ {\bf a}_z \nabla_{t,\hat{r}} \cdot \left[ {\bf a}_z \times \int_{S_4} {\bf{h}}^{0} \, dS \right] \,\, . \\
\end{array}
\label{bch224f}
\end{equation}
The last terms on the right side of (\ref{bch2}) can be transformed by using the portion of Amp\`{e}re's Law transverse to $y$:
\begin{equation}
{\bf a}_y\times\left. \frac{\partial \bf{H}^0}{\partial {\hat y}}\right|_{{\bf r}_o} = j\frac{\epsilon_r}{\eta_0} \mathbf{E}_t^0({\bf r}_o)
+\frac{1}{\mu_r} {\bf a}_y\times\nabla_{t,\hat r} \left[ \frac{{{B}}^0_y({\bf{r}}_0)}{\mu_0} \right]
\label{fe3b}
\end{equation}
Combining (\ref{bch2})-(\ref{bch224f}), and using (\ref{dsce}) and (\ref{dsch}), we obtain
\begin{equation}
\begin{array}{c}
{\bf{a}}_{y}\times\left[{\bf{H}}^{\rmA 1}({\bf r}_o)-{\bf{H}}^{\rmB 1}({\bf r}_o)\right]= \\
  {\bf{a}}_x\,\frac{j}{\eta_{0}}\,
\left[\frac{\hat{\chi}_{ES}^{xx}}{p} {E}^0_{x}({\mathbf{r}}_o) + \frac{\hat{\chi}_{ES}^{xy}}{p} \frac{{D}^0_{y}({\mathbf{r}}_o)}{\epsilon_0} + \frac{\hat{\chi}_{ES}^{xz}}{p} {E}^0_{z}({\mathbf{r}}_o) \right] \\
\mbox{}+ {\bf{a}}_z\,\frac{j}{\eta_{0}}\,
\left[\frac{\hat{\chi}_{ES}^{zx}}{p} {E}^0_{x}({\mathbf{r}}_o) + \frac{\hat{\chi}_{ES}^{zy}}{p} \frac{{D}^0_{y}({\mathbf{r}}_o)}{\epsilon_0} + \frac{\hat{\chi}_{ES}^{zz}}{p} {E}^0_{z}({\mathbf{r}}_o) \right] \\
\mbox{}- {\bf{a}}_y\times \nabla_{t,\hat r} \left[ \frac{\hat{\chi}_{MS}^{yx}}{p} {H}^{0}_{x} + \frac{\hat{\chi}_{MS}^{yy}}{p} \frac{B^0_{y}(\mathbf{r}_o)}{\mu_0} + \frac{\hat{\chi}_{MS}^{yz}}{p} {H}^{0}_{z}({\mathbf{r}}_o) \right]\,\, .\\
\end{array}
\label{bch2tb}
\end{equation}
These remaining effective surface susceptibilities dyadics are given in (\ref{chiEE}) and (\ref{chiMM}).  The expressions (\ref{chiEE}) and (\ref{chiMM}), are the components of the $3 \times 3$ dyadic electric and magnetic surface susceptibilities needed to fully characterize the metafilm.

Now that we have boundary conditions for the zeroth-order and
first-order fields [i.~e., equations (\ref{bceo1}), (\ref{bcho1}), (\ref{bce1effta}),
 and (\ref{bch2tb})], boundary conditions for the total effective fields can be obtained.  Using
equation (\ref{ehsp}), the boundary condition for the total
effective $E$-field at the $y=0$ plane is expressed to first order
in $\nu$ as
\begin{equation}
\begin{array}{c}
{\bf{a}}_{y}\times\left[{\bf{E}}^{\rmA}({\mathbf{r}}_o)-{\bf{E}}^{\rmB}({\mathbf{r}}_o) \right]=
 {\bf{a}}_{y}\times\left[{\bf{E}}^{\rmA0}({\mathbf{r}}_o)-{\bf{E}}^{\rmB0 }({\mathbf{r}}_o)\right] \\
+\nu\,\,{\bf{a}}_{y}\times\left[{\bf{E}}^{\rmA1}({\mathbf{r}}_o)-{\bf{E}}^{\rmB 1}({\mathbf{r}}_o) \right]+O(\nu^{2}) \,\,\, .
\end{array}
\end{equation}
By (\ref{bceo1}), the first term of the
right side of this expression is zero. Using $\nu=pk_{o}$, $\frac{\partial}{\partial
{\hat x}} = \frac{1}{k_{o}} \frac{\partial}{\partial x}$, and $\frac{\partial}{\partial
{\hat z}} = \frac{1}{k_{o}} \frac{\partial}{\partial z}$ the
boundary condition for the effective $E$-field can be written in terms of the original unscaled variables as
\begin{equation}
\begin{array}{c}
{\bf{a}}_{y}\times
     \left[\mathbf{E}_A-\mathbf{E}_{B}\right]=\\
      - j\omega\mu_0 p \left\{
{\bf{a}}_x\left[ \hat{\chi}_{MS}^{xx}
{H}^0_{x}({\mathbf{r}}_o) + \hat{\chi}_{MS}^{xy}
\frac{{B}^0_{y}({\mathbf{r}}_o)}{\mu_0}
+ \hat{\chi}_{MS}^{xz} {H}^0_{z}({\mathbf{r}}_o)\right]\right. \\
\left. + {\bf{a}}_z \left[\hat{\chi}_{MS}^{zx}
{H}^0_{x}({\mathbf{r}}_o) + \hat{\chi}_{MS}^{zy}
\frac{{B}^0_{y}({\mathbf{r}}_o)}{\mu_0}
+ \hat{\chi}_{MS}^{zz} {H}^0_{z}({\mathbf{r}}_o)\right] \right\}\\
   -p {\bf{a}}_y\times \nabla_{t} \left[ \hat{\chi}_{ES}^{yx}
{E}^{0}_{x}+ \hat{\chi}_{ES}^{yy}
\frac{D^0_{y}(\mathbf{r}_o)}{\epsilon_0} +
\hat{\chi}_{ES}^{yz} {E}^{0}_{z}\right] \, ,\\
\end{array}
\label{tebc}
\end{equation}
and in a similar way,
\begin{equation}
\begin{array}{c}
{\bf{a}}_{y}\times
     \left[\mathbf{H}_A-\mathbf{H}_{B}\right]=\\
     j\omega\epsilon_0 p \left\{
{\bf{a}}_x \left[\hat{\chi}_{ES}^{xx}
{E}^0_{x}({\mathbf{r}}_o) + \hat{\chi}_{ES}^{xy}
\frac{{D}^0_{y}}{\epsilon_0}({\mathbf{r}}_o)
+ \hat{\chi}_{ES}^{xz} {E}^0_{z}({\mathbf{r}}_o) \right] \right. \\
\left.+{\bf{a}}_z \left[\hat{\chi}_{ES}^{zx}
{E}^0_{x}({\mathbf{r}}_o) + \hat{\chi}_{ES}^{zy}
\frac{{D}^0_{y}}{\epsilon_0}({\mathbf{r}}_o)
+ \hat{\chi}_{ES}^{zz} {E}^0_{z}({\mathbf{r}}_o)\right] \right\}\\
-p {\bf{a}}_y\times \nabla_{t} \left[ \hat{\chi}_{MS}^{yx}
{H}^{0}_{x}+ \hat{\chi}_{MS}^{yy}
\frac{B^0_{y}(\mathbf{r}_o)}{\mu_0} +
\hat{\chi}_{MS}^{yz} {H}^{0}_{z} \right] \, ,\\
     \end{array}
\label{thhbca}
\end{equation}

Although the zeroth-order fields (${E}^0_{x}$, ${D}^0_{y}$, ${E}^0_{z}$, ${H}^0_{x}$, ${B}^0_{y}$, and ${H}^0_{z}$) appearing in the right sides of these expressions are continuous across the interface, the same is not true of terms of higher order ($m \geq 1$), so there remains some ambiguity about how to express these right sides in terms of the total effective fields $\mathbf{E}^{\rmA}$, $\mathbf{E}^{\rmB}$, etc. It can be shown (the details will not be given here, but are analogous to the derivations done in \cite{pere1}-\cite{pere2}) that if we replace the fields $\mathbf{E}^0$, $\mathbf{H}^0$, etc. by the average fields at the interface as in (\ref{eavv}),
\begin{equation}
\mathbf{E}_{\rm av}=\frac{1}{2} \left( \mathbf{E}^{\rmA}+\mathbf{E}^{\rmB} \right) \,\,\, ,
\label{avge}
\end{equation}
and similarly for $\mathbf{H}_{\rm av}$, $\mathbf{D}_{\rm av}$, and $\mathbf{B}_{\rm av}$, the resulting boundary conditions are still correct to the same order [$O(k_0^2 p^2)$], but will satisfy reciprocity and conservation of energy exactly. This modification will ensure that numerical or analytical difficulties will not arise when these boundary conditions are employed. Moreover, use of this symmetric average has been shown to produce greater accuracy in numerical simulations \cite{duru} (see also \cite{chun}). Thus, the final forms of the jump conditions on the tangential effective fields are:
\begin{eqnarray}
\lefteqn{{\bf{a}}_{y}\times \left[\mathbf{E}^{\rmA}-\mathbf{E}^{\rmB}\right]_{y=0} =} \label{tebct} \\
  && - j\omega\mu_0 \left( \dyadictall{\boldsymbol{\chi}}_{MS} \cdot \tilde{\mathbf{H}}_{\rm av} \right)_t - \mathbf{a}_y \times \nabla_{t} \left( \mathbf{a}_y \cdot \dyadictall{\boldsymbol{\chi}}_{ES} \cdot \tilde{\mathbf{E}}_{\rm av} \right) \nonumber
\end{eqnarray}
and
\begin{eqnarray}
\lefteqn{{\bf{a}}_{y}\times \left[\mathbf{H}^{\rmA}-\mathbf{H}^{\rmB}\right]_{y=0} =} \label{thhbc} \\
  && j\omega\epsilon_0 \left( \dyadictall{\boldsymbol{\chi}}_{ES} \cdot \tilde{\mathbf{E}}_{\rm av} \right)_t - \mathbf{a}_y \times \nabla_{t} \left( \mathbf{a}_y \cdot \dyadictall{\boldsymbol{\chi}}_{MS} \cdot \tilde{\mathbf{H}}_{\rm av} \right) \nonumber
\end{eqnarray}
where we have used the notations
\begin{equation}
 \tilde{\mathbf{E}}_{\rm av} = \mathbf{a}_x {E}_{{\rm av},x}({\mathbf{r}}_o) + \mathbf{a}_y \frac{{D}_{{\rm av},y}({\mathbf{r}}_o)}{\epsilon_0} + \mathbf{a}_z {E}_{{\rm av},z}({\mathbf{r}}_o)
\end{equation}
\begin{equation}
 \tilde{\mathbf{H}}_{\rm av} = \mathbf{a}_x {H}_{{\rm av},x}({\mathbf{r}}_o) + \mathbf{a}_y \frac{{B}_{{\rm av},y}({\mathbf{r}}_o)}{\mu_0} + \mathbf{a}_z {H}_{{\rm av},z}({\mathbf{r}}_o)
\end{equation}
and the surface susceptibility dyadics are defined as
\begin{eqnarray}
\dyadic{\boldsymbol{\chi}}_{ES} & = & \chi_{ES}^{xx} {\bf a}_x {\bf a}_x +
\chi_{ES}^{xy} {\bf a}_x {\bf a}_y + \chi_{ES}^{xz} {\bf a}_x {\bf a}_z \label{chigene} \\
&& + \chi_{ES}^{yx} {\bf a}_y {\bf a}_x +
\chi_{ES}^{yy} {\bf a}_y {\bf a}_y + \chi_{ES}^{yz} {\bf a}_y {\bf a}_z \nonumber \\
&& +  \chi_{ES}^{zx} {\bf a}_x {\bf a}_x +
\chi_{ES}^{zy} {\bf a}_z {\bf a}_y + \chi_{ES}^{zz} {\bf a}_z {\bf a}_z \nonumber
\end{eqnarray}
\begin{eqnarray}
\dyadic{\boldsymbol{\chi}}_{MS} & = & \chi_{MS}^{xx} {\bf a}_x {\bf a}_x +
\chi_{MS}^{xy} {\bf a}_x {\bf a}_y + \chi_{MS}^{xz} {\bf a}_x {\bf a}_z \label{chigenm} \\
&& + \chi_{MS}^{yx} {\bf a}_y {\bf a}_x +
\chi_{MS}^{yy} {\bf a}_y {\bf a}_y + \chi_{MS}^{yz} {\bf a}_y {\bf a}_z \nonumber \\
&& + \chi_{MS}^{zx} {\bf a}_x {\bf a}_x +
\chi_{MS}^{zy} {\bf a}_z {\bf a}_y + \chi_{MS}^{zz} {\bf a}_z {\bf a}_z \nonumber \,\, .
\end{eqnarray}

The GSTCs (\ref{tebct}) and (\ref{thhbc}) are the main results of this paper, and we see that they have the same basic functional form as eqns.~(\ref{bce}) above, which were derived in \cite{kmh} using an approach based on the approximation of only dipole interaction of the scatterers. One difference from equations (\ref{bce}) is that the material parameters $\epsilon_{A,B}$ and $\mu_{A,B}$ of the half-spaces on either side of the metafilm are now embedded in the definitions of the susceptibilities $\dyadictall{\boldsymbol{\chi}}_{ES}$ and $\dyadictall{\boldsymbol{\chi}}_{MS}$ rather than being displayed (less appropriately) as explicit factors in the GSTCs. Another difference between (\ref{tebct})-(\ref{thhbc}) and (\ref{bce}) is that our new expressions can have off-diagonal terms in both the electric and magnetic surface susceptibilities.  These expressions show that full dyadic surface susceptibilities (including off-diagonal elements) are needed to fully characterize a metafilm composed of arbitrarily-shaped scatterers. That these off-diagonal terms should be different from zero was conjectured in \cite{calzo}, but no proof was given. The results in \cite{calzo} show the importance that these off-diagonal terms can have in the reflection and transmission at a metafilm. Off-diagonal terms have also been found to be generally present in the GSTCs derived for an arbitrarily-shaped, material-coated wire grating \cite{wirehk}. We may finally remark that our homogenization approach does not require some of the assumptions and approximations inherent in the dipole interaction approach---in particular, we can allow the scatterers to be closely packed.

\section{Comparisons to the Dipole Approximation}

To compare the results of this paper to those of \cite{kmh}, which are based on dipole interactions only and thus limited to sparsely spaced scatterers, we investigate an array of perfectly conducting spheres. To determine the susceptibilities as derived in the present paper, solutions of the boundary problems for the normalized boundary-layer fields given in Appendix~\ref{apa} are required, and then  various integrals of these fields must be carried out as described in Appendix~\ref{aplast}.  We used the commercial numerical program COMSOL (mention of this software is not an endorsement but is only intended to
clarify what was done in this work) to numerically solve these static boundary problems and to evaluate the various integrals for the case of the sphere array.

The surface susceptibilities obtained from the dipole approach are given in (17)-(22) of \cite{hk2}, where they are expressed in terms of the electric and magnetic polarizabilities of the spheres. When using (17)-(22) in \cite{hk2}, note that the normal direction to the interface in \cite{hk2} is $z$, while in this paper the normal direction has been taken as $y$.  For a perfect conducting sphere, the electric polarizability is $3V$ and the magnetic polarizability is $-3V/2$, where $V$ is the volume of the sphere. Fig.~\ref{compare} shows the calculated values for $\chi_{ES}^{yy}/p$, $\chi_{ES}^{xx}/p$, $\chi_{MS}^{yy}/p$, and $\chi_{MS}^{xx}/p$ as functions of the sphere radius ($a$) normalized to the period ($p$). The susceptibilities from the dipole approach are also shown for comparison.  We see good agreement between the numerically calculated values and the dipole-interaction results when $a/p < 0.25$, but beyond that filling density, the dipole approach breaks down and is inaccurate for closely packed scatterers. The multiple-scale approach presented here does not have this limitation. Indeed, the multiple-scale results show that the values of $\chi_{ES}^{xx}$ and $\chi_{MS}^{yy}$ become very large as $a/p\rightarrow0.5$. It is known that the effective permittivity of a three-dimensional array of spheres becomes infinite in the limit as the spheres touch \cite{keller}-\cite{andrian}; it seems likely that a similar assertion is true for these surface susceptibilities of the metafilm.

A further justification of this multiple-scale homogenization approach for these GSTCs was given in \cite{wirehk} where we compared the surface susceptibility for a two-dimensional wire-grating to those obtained from a different approach. The surface susceptibilities obtained for the two-dimensional wire-grating are analogous to those obtained from the three-dimensional approach given here. In fact, in \cite{wirehk}, it is shown that the term for the wire-grating that is equivalent to that of $\chi_{ES}^{xx}/p$ for the metafilm also becomes very large as $a/p\rightarrow0.5$, see Fig. 7 in \cite{wirehk}.

\begin{figure}
\centering
\scalebox{0.35} {\includegraphics*{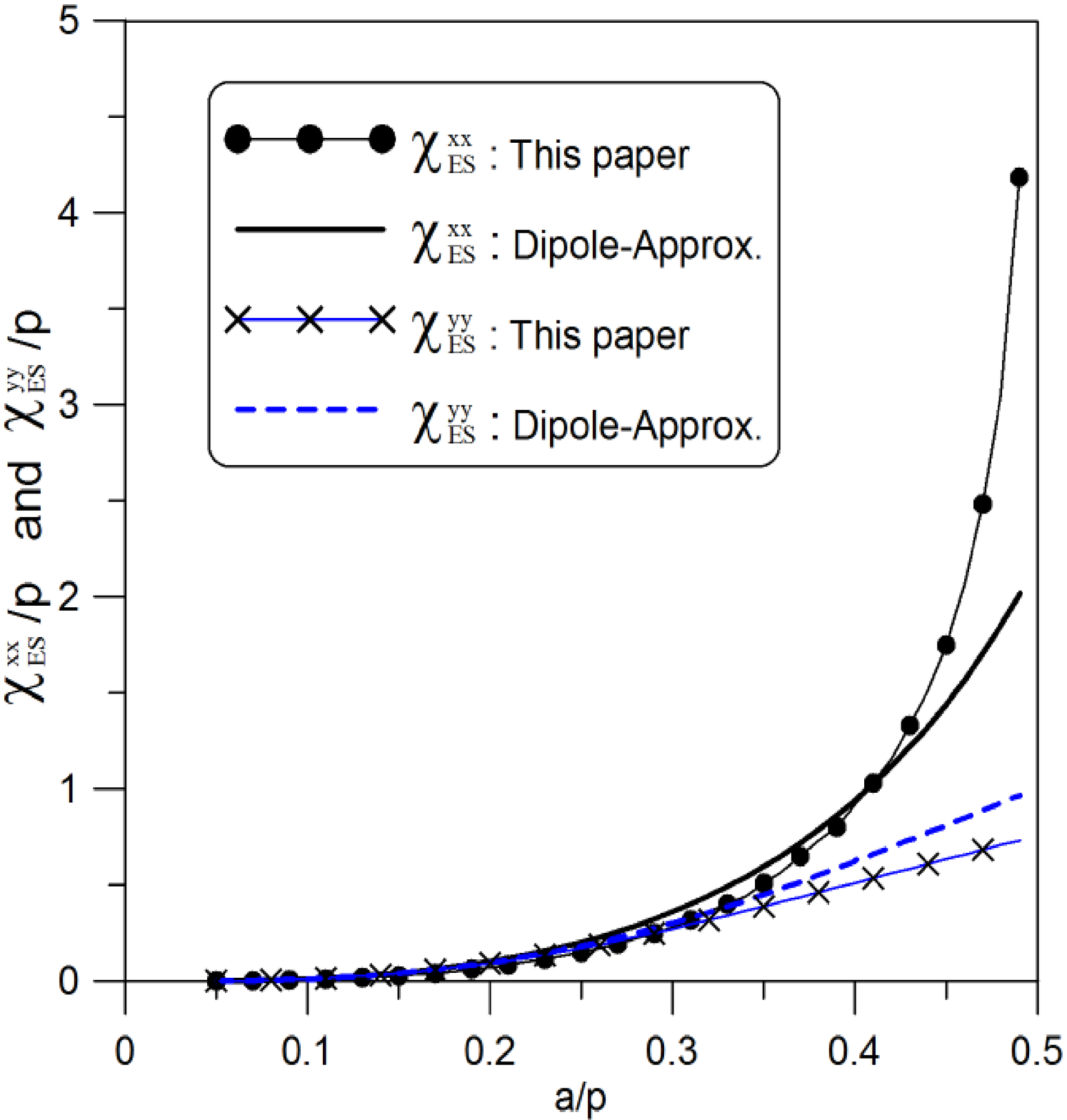}}
\begin{center}
\footnotesize(a)
\end{center}
\centering
\scalebox{0.35} {\includegraphics*{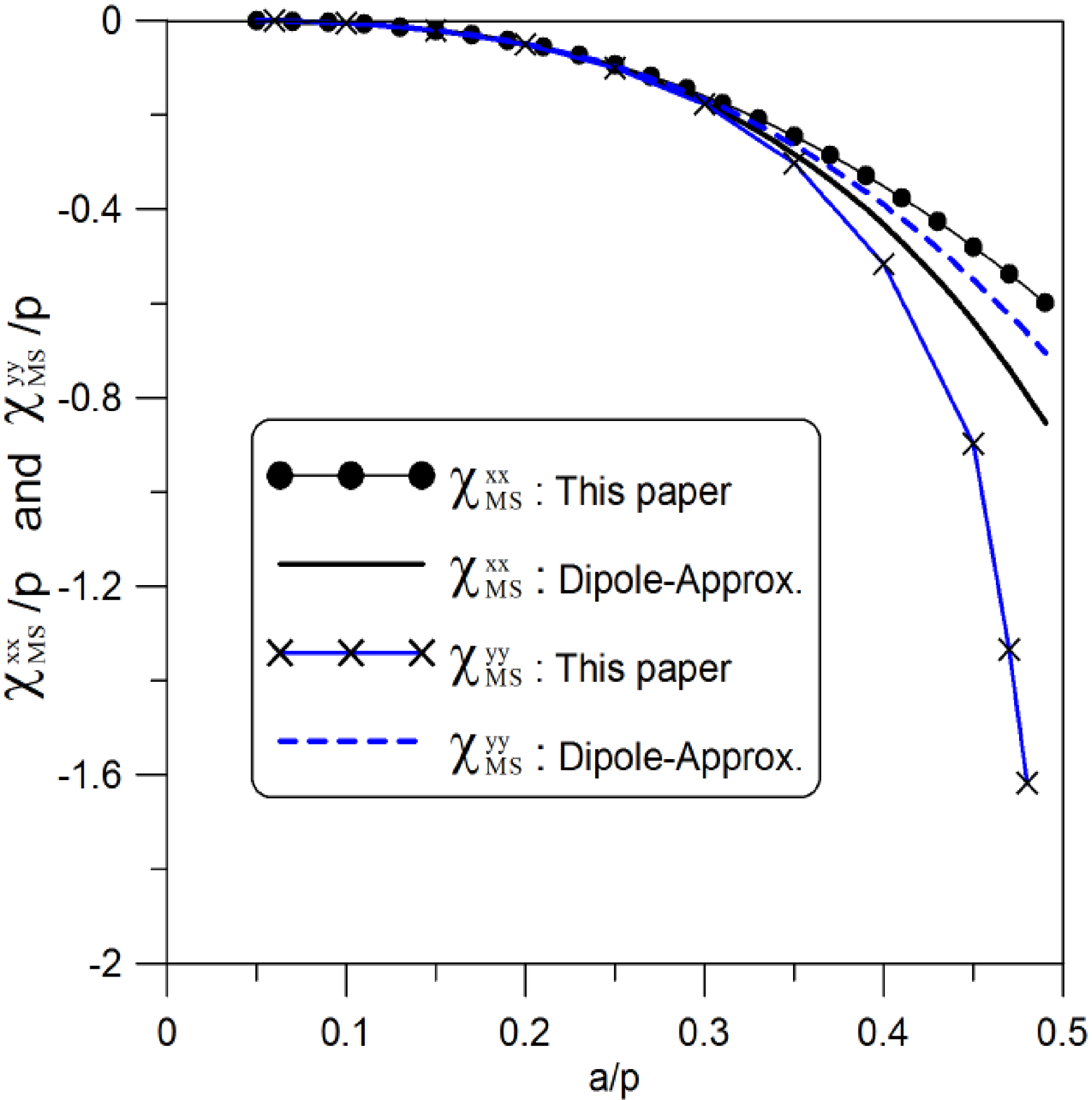}}
\begin{center}
\footnotesize(b)
\end{center}
\caption{Comparison of surface susceptibilities for an array of perfectly conducting spheres: (a) $\chi_{ES}^{yy}/p$ and $\chi_{ES}^{xx}/p$,  (b) $\chi_{MS}^{yy}/p$ and $\chi_{MS}^{xx}/p$.} \label{compare}
\end{figure}
\normalsize

\section{Conclusion and Discussion}
\label{s5}

We have shown how a multiple-scale homogenization method can be used to derive GSTCs for electromagnetic fields on the surface of a metafilm.
The parameters in these boundary conditions are effective electric and magnetic surface susceptibilities, which are related to the geometry of the scatterers that constitute the composite.  We have shown that full dyadic surface susceptibilities are needed to fully characterize a metafilm composed of generic, arbitrarily-shaped scatterers.

While in this paper we have considered only the case of PEC scatterers, a similar but more involved derivation can be carried out for the case of non-PEC scatterers, but the final form of the desired GSTCs will be the same. In examining how this work might be further extended, we have shown that expressions for the surface susceptibilities can be even more complicated, and exhibit very interesting properties such as bianisotropy. Bianisotropy can also arise when a metafilm is located near a material interface \cite{powell}-\cite{alboo2}, but it can be shown that this effect is of a higher order, $O(\nu^2)$, and therefore does not appear at the order of approximation reached in the present paper. Also not covered by our results here is the effect of resonance in the scatterers. To handle this would require modification of our technique to what is sometimes called ``stiff'' homogenization; examples of this can be found in \cite{bouch1}-\cite{bouch4}.

Using the homogenization technique, we have laid out a framework for the calculation of the surface susceptibilities, which requires the solution of a set of static field problems. As illustrated by the example of section IV, calculating these static fields and surface susceptibilities will in general have to be done by numerical means.  However, the GSTCs, as derived here, can be used as the basis of a technique to retrieve these surface parameters from measured or computed reflection and transmission data, the results then used in applications to analyze various problems of interest.  This is analogous to what was done in characterizing metasurfaces and interface problems in \cite{hk3}, \cite{hk2}, \cite{awpl} and \cite{hr7}.

Finally, we have extended the work presented here by adapting it and combining it with ideas developed for wire gratings \cite{wirehk} to derive a set of GSTCs for metascreens. This will be the topic of another publication.

\appendices

\section{Geometric Integrals and Other Identities}
\label{ap0}

We collect here several integrals whose values depend only on the geometry of the scatterer. We have the elementary result
\begin{equation}
 \int_{\partial A_g/\partial B_g}\,dS = \hat{S}_g
\label{lgap}
\end{equation}
where
\begin{equation}
{\hat{S}}_g=1-{\hat{S}}_p
\label{defineg}
\end{equation}
is the area of the gap region intersected by the plane $\xi_y=0$, and ${\hat{S}_p}$ is the area of the cross section of the scatterer intersected by the plane $\xi_y=0$, both in scaled dimensions (the actual areas are $S_p={\hat{S}_p}p^2$ and $S_g={\hat{S}_g}p^2$). The next identities follow from the divergence theorem. First,
\begin{equation}
\begin{array}{c}
\int_{\partial A_s}{\bf{a}}_{n}\,dS= {\bf{a}}_{y}\,{\hat{S}_p}\,\,\,\, {\rm and}\,\,\,\,
\int_{\partial B_s}{\bf{a}}_{n}\,dS= -{\bf{a}}_{y}\,{\hat{S}_p}\,\,\, .\\
\end{array}
\label{temp1}
\end{equation}
Second,
\begin{equation}
\begin{aligned}
\int_{\partial A_s} \xi_{k} {\bf{a}}_n \,dS &= \hat{V}_{sA} \mathbf{a}_{k} + \hat{S}_p \xi_{pk} \mathbf{a}_y \\
\int_{\partial B_s} \xi_{k} {\bf{a}}_n \,dS &= \hat{V}_{sB} \mathbf{a}_{k} - \hat{S}_p \xi_{pk} \mathbf{a}_y
\end{aligned}
\label{Vwa}
\end{equation}
for $k = x, y$ or $z$, where $\hat{V}_{sA}$ and $\hat{V}_{sB}$ are the scaled volumes of the scatterer that are above and below the $\xi_y=0$ reference plane respectively (so that $\hat{V}_s=\hat{V}_{sA}+\hat{V}_{sB}$), and $\boldsymbol{\xi}_p = \xi_{px} \mathbf{a}_x + \xi_{pz} \mathbf{a}_z$ is the centroid of $S_p$ (note that $\xi_{py} = 0$):
\begin{equation}
 \boldsymbol{\xi}_p = \left. \frac{1}{\hat{S}_p} \int_{S_p} \boldsymbol{\xi} \, dS \right|_{\xi_y = 0}
\end{equation}

We will need two further identities that were presented in equations (145) and (151) of \cite{hr7}. If $\mathbf{F}(\boldsymbol{\xi})$ is any vector function whose tangential components are continuous on a closed surface $S$, then
\begin{equation}
 \oint_{S} \mathbf{a}_n \cdot \nabla_{\xi} \times \mathbf{F} \, dS = 0
 \label{curlid}
\end{equation}
and
\begin{equation}
 \oint_{S} \mathbf{a}_n \times \mathbf{F} \, dS = \oint_{S} \boldsymbol{\xi} \mathbf{a}_n \cdot \nabla_{\xi} \times \mathbf{F} \, dS
 \label{klein}
\end{equation}
Two final identities are also useful, that are proved by elementary means:
\begin{equation}
 \mathbf{A}_1 \cdot \left[ \mathbf{A}_2 \times \left( \mathbf{A}_3 \times \mathbf{A}_4 \right) \right] = - \mathbf{A}_3 \cdot \left[ \left( \mathbf{A}_1 \times \mathbf{A}_2\right)  \times \mathbf{A}_4 \right]
 \label{a14}
\end{equation}
for any vectors $\mathbf{A}_1, \ldots , \mathbf{A}_4$, and
\begin{equation}
 \nabla_{\xi} \cdot \left[ \left( \mathbf{a}_i \times \boldsymbol{\xi} \right) \times \mathbf{F} \right] = 2 \mathbf{a}_i \cdot \mathbf{F}
 \label{3pai}
\end{equation}
for any vector function $\mathbf{F}$ that obeys $\nabla_{\xi} \times \mathbf{F} = 0$, where $\mathbf{a}_i = \mathbf{a}_x, \mathbf{a}_y$ or $\mathbf{a}_z$.

\section{Integral Constraints (Solvability Conditions) for the Boundary-Layer Fields}
\label{apbl}

Stokes' theorem can be applied to the curl of ${\bf{e}}^m$ by  integrating it over the volume $V_A$ shown in Figs. \ref{fig3} and \ref{fig4} to give
\begin{equation}
\int_{V_A} \nabla_{\xi}\times{\bf{e}}^{m} \,\, dV = -\oint_{\partial A} \mathbf{a}_n \times {\bf{e}}^{m} \, dS
\label{curleo2}
\end{equation}
The integral over the boundary of $V_A$ breaks up into
\begin{equation}
\oint_{\partial A}=\int_{\partial A_g}+\int_{\partial A_s}+\int_{\partial A_{\infty}} + \sum_{n=1}^{4}\int_{\partial A_n}\,\, ,
\label{sureo}
\end{equation}
where $\partial A_n$ represent the four vertical sides of $V_A$.
Due to periodicity, the integrals over the four sides ($\sum\int_{\partial A_n}$)
cancel, and because ${\bf{e}}^{m} \rightarrow 0$ as
$|\xi_y|\rightarrow\infty$, the third term on the right side of equation
(\ref{sureo}) vanishes.  Thus, equation (\ref{curleo2}) reduces to
\begin{equation}
{\bf{a}}_y\times\int_{\partial A_g}{\bf{e}}^{\rmA m} \,
dS +\int_{\partial A_s}{\bf{a}}_n\times{\bf{e}}^{\rmA m} \,
dS= -\int_{V_A} \nabla_{\xi}\times{\bf{e}}^{m}\, dV \,\, .
\label{sureo3}
\end{equation}
In a similar manner, we carry out an integral of $\nabla_{\xi}\times{\bf{e}}^{m}$ over the volume $V_B$ shown in Figs. \ref{fig3} and \ref{fig4}. With the indicated directions of the surface normals ${\bf a}_n$, we find
\begin{equation}
-{\bf{a}}_y\times\int_{\partial B_g}{\bf{e}}^{\rmB m} \, dS +\int_{\partial B_s}{\bf{a}}_n\times{\bf{e}}^{\rmB m} \, dS= -\int_{V_B} \nabla_{\xi}\times{\bf{e}}^{m} \, dV
\label{sureo3a}
\end{equation}
having used the fact that ${\bf{a}}_n=-{\bf{a}}_y$ on $\partial B_g$. By adding (\ref{sureo3a}) to (\ref{sureo3}) we obtain
\begin{equation}
\begin{array}{c}
{\bf{a}}_y\times\int_{\partial A_g/\partial B_g}\left[{\bf{e}}^{{\rm A}m} -{\bf{e}}^{{\rm B}m}\right]\,dS
+\oint_{S_s}{\bf{a}}_n\times{\bf{e}}^{m} \,dS
= \\-\int_{V} \nabla_{\xi}\times{\bf{e}}^{m}\, dV \,\,\, .
\end{array}
\label{eointeg1}
\end{equation}
Finally, using (\ref{lgap}) and the boundary condition (\ref{dobc3}) in the gap, we have
\begin{eqnarray}
 \lefteqn{\hat{S}_g \mathbf{a}_y \times \left[{\bf{E}}^{\rmA m}({\bf r}_o) -{\bf{E}}^{\rmB m}({\bf r}_o)\right] =} \label{solvem} \\
 && \oint_{S_s} \mathbf{a}_n \times \mathbf{e}^m \, dS + \int_V \nabla_{\xi} \times \mathbf{e}^m \, dV \nonumber
\end{eqnarray}
which is a solvability condition for the boundary-layer field $\mathbf{e}^m$. An exactly similar derivation using the boundary condition (\ref{ht0bc}) in the gap leads to a solvability condition for $\mathbf{h}^m$:
\begin{eqnarray}
 \lefteqn{\hat{S}_g \mathbf{a}_y \times \left[{\bf{H}}^{\rmA m}({\bf r}_o) -{\bf{H}}^{\rmB m}({\bf r}_o)\right] =} \label{solvhm} \\
 && \oint_{S_s} \mathbf{a}_n \times \mathbf{h}^m \, dS + \int_V \nabla_{\xi} \times \mathbf{h}^m \, dV \nonumber
\end{eqnarray}
In an analogous way, we can obtain solvability conditions by use of the divergence theorem on $\nabla_{\xi} \cdot ( \epsilon_r \mathbf{e}^m)$ and $\nabla_{\xi} \cdot ( \mu_r \mathbf{h}^m)$, together with the gap boundary conditions (\ref{dn0bc}) and (\ref{dobc3b}). We have for $\mathbf{d}^m$:
\begin{eqnarray}
 \lefteqn{\hat{S}_g \mathbf{a}_y \cdot \left[{\bf{D}}^{\rmA m}({\bf r}_o) -{\bf{D}}^{\rmB m}({\bf r}_o)\right] =} \label{solvdm} \\
 && \epsilon_0 \left[ \oint_{S_s} \epsilon_r \mathbf{a}_n \cdot \mathbf{e}^m \, dS + \int_V \nabla_{\xi} \cdot \left( \epsilon_r \mathbf{e}^m \right) \, dV \right] \nonumber
\end{eqnarray}
and for $\mathbf{b}^m$:
\begin{eqnarray}
 \lefteqn{\hat{S}_g \mathbf{a}_y \cdot \left[{\bf{B}}^{\rmA m}({\bf r}_o) -{\bf{B}}^{\rmB m}({\bf r}_o)\right] =} \label{solvbm} \\
 && \mu_0 \left[ \oint_{S_s} \mu_r \mathbf{a}_n \cdot \mathbf{h}^m \, dS + \int_V \nabla_{\xi} \cdot \left( \mu_r \mathbf{h}^m \right) \, dV \right] \nonumber
\end{eqnarray}

We can make further progress in the reduction of the solvability conditions (\ref{solvem}) and (\ref{solvbm}) by employing the boundary conditions (\ref{dobc})-(\ref{d1bc2}) together with (\ref{defineg}), (\ref{temp1}) and (\ref{Vwa}) to express the integrals over $S_s$ in terms of the macroscopic fields. For $m=0$ we have
\begin{equation}
 \mathbf{a}_y \times \left[{\bf{E}}^{\rmA 0}({\bf r}_o) - {\bf{E}}^{\rmB 0}({\bf r}_o)\right] = \int_V \nabla_{\xi} \times \mathbf{e}^0 \, dV
 \label{solvem0}
\end{equation}
\begin{equation}
 \mathbf{a}_y \cdot \left[{\bf{B}}^{\rmA 0}({\bf r}_o) - {\bf{B}}^{\rmB 0}({\bf r}_o)\right] = \mu_0 \int_V \nabla_{\xi} \cdot \left( \mu_r \mathbf{h}^0 \right) \, dV
 \label{solvbm0}
\end{equation}
and for $m=1$,
\begin{eqnarray}
 \lefteqn{\mathbf{a}_y \times \left[{\bf{E}}^{\rmA 1}({\bf r}_o) - {\bf{E}}^{\rmB 1}({\bf r}_o)\right] = \int_V \nabla_{\xi} \times \mathbf{e}^1 \, dV} \nonumber \\
 && - \mathbf{a}_y \times \left[ \hat{V}_{sA} \frac{\partial \mathbf{E}^{\rmA 1}}{\partial \hat{y}} + \hat{V}_{sB} \frac{\partial \mathbf{E}^{\rmB 1}}{\partial \hat{y}} \right]_{y=0} \label{solvem1}
\end{eqnarray}
\begin{eqnarray}
 \lefteqn{\mathbf{a}_y \cdot \left[{\bf{B}}^{\rmA 1}({\bf r}_o) - {\bf{B}}^{\rmB 1}({\bf r}_o)\right] = \mu_0 \int_V \nabla_{\xi} \cdot \left( \mu_r \mathbf{h}^1 \right) \, dV} \nonumber \\
 && - \mathbf{a}_y \cdot \left[ \hat{V}_{sA} \frac{\partial \mathbf{B}^{\rmA 1}}{\partial \hat{y}} + \hat{V}_{sB} \frac{\partial \mathbf{B}^{\rmB 1}}{\partial \hat{y}} \right]_{y=0} \label{solvbm1}
\end{eqnarray}
However, there are no analogous boundary conditions for normal $\mathbf{d}^m$ or tangential $\mathbf{h}^m$ on $S_s$, so a different approach must be used.

As in \cite{hr7}, we can use (\ref{curlid}) to show that
\begin{equation}
\oint_{S_s} \epsilon\,{\bf{a}}_n \cdot {\bf{E}}^T \, dS = 0 \label{ai3}
\end{equation}
This states that the total surface charge on each scatterer is zero. Now the integrand of (\ref{ai3}) can be expanded in powers of $\nu$ using (\ref{ehsp}) and (\ref{taylor}). If we take only terms of order $\nu^0$ in this equation, we get
\begin{equation}
\oint_{S_s}\epsilon\,{\bf{a}}_n \cdot {\bf{e}}^0 \,\,dS=-{\bf{a}}_{y}\cdot\left[{\bf{D}}^{\rmA 0}-{\bf{D}}^{\rmB 0}\right]\,{\hat{S}_p}\,\, ,
\label{ine0}
\end{equation}
whereas if we take terms of order $\nu^1$, we get
\begin{eqnarray}
\oint_{S_s}\epsilon\,{\bf{a}}_n \cdot {\bf{e}}^1 \,\,dS &=& -{\bf{a}}_{y}\cdot\left[{\bf{D}}^{\rmA 1}-{\bf{D}}^{\rmB 1}\right]\,{\hat{S}_p} \label{ine1} \\
&& - {\bf{a}}_{y}\cdot\left[ \hat{V}_{sA} \frac{\partial {\bf{D}}^{\rmA 0}}{\partial \hat{y}} + \hat{V}_{sB} \frac{\partial {\bf{D}}^{\rmB 0}}{\partial \hat{y}} \right]_{y=0} \,\, ,
\nonumber
\end{eqnarray}
since ${\bf{D}}^{(A,B)}$ are independent of $\bfxi$. Therefore from (\ref{solvdm}) we obtain solvability conditions for $m=0$:
\begin{equation}
 \mathbf{a}_y \cdot \left[{\bf{D}}^{\rmA 0}({\bf r}_o) -{\bf{D}}^{\rmB 0}({\bf r}_o)\right] = \epsilon_0 \int_V \nabla_{\xi} \cdot \left( \epsilon_r \mathbf{e}^0 \right) \, dV \label{solvdm0}
\end{equation}
and for $m=1$:
\begin{eqnarray}
 \lefteqn{\mathbf{a}_y \cdot \left[{\bf{D}}^{\rmA 1}({\bf r}_o) -{\bf{D}}^{\rmB 1}({\bf r}_o)\right] = \epsilon_0 \int_V \nabla_{\xi} \cdot \left( \epsilon_r \mathbf{e}^0 \right) \, dV} \nonumber \\
 && - {\bf{a}}_{y}\cdot\left[ \hat{V}_{sA} \frac{\partial {\bf{D}}^{\rmA 0}}{\partial \hat{y}} + \hat{V}_{sB} \frac{\partial {\bf{D}}^{\rmB 0}}{\partial \hat{y}} \right]_{y=0} \label{solvdm1}
\end{eqnarray}
An analogous result for $\mathbf{h}^m$ is achieved starting by taking $\mathbf{F} = {\bf h} + {\bf H}$ in identity (\ref{klein}) to obtain
\begin{equation}
\oint_{S_s} {\bf a}_n \times {\bf h}\, dS+\oint_{S_s} {\bf a}_n \times {\bf H} \, dS = \oint_{S_s} \bfxi {\bf a}_n \cdot \nabla_{\xi} \times {\bf h} \, dS
\label{aiq2}
\end{equation}
because $\nabla_{\xi} \times {\bf H} = 0$. Once again, the integrands in (\ref{aiq2}) can be expanded using (\ref{ehsp}) and (\ref{taylor}), so grouping terms of order $\nu^0$ and $\nu^1$ separately and using (\ref{temp1}) and (\ref{Vwa}), we obtain at orders $m=0$ and $m=1$:
\begin{eqnarray}
\oint_{S_s} {\bf{a}}_{n}\times{\bf{h}}^{0} \,dS & = & -{\bf{a}}_{y}\times\left[{\bf{H}}^{\rmA 0}-{\bf{H}}^{\rmB 0}\right]\,{\hat{S}_p} \nonumber \\
&& + \oint_{S_s} \bfxi {\bf a}_n \cdot \nabla_{\xi} \times {\bf h}^0 \, dS \,\,\, ,
\label{bcscatf3A}
\end{eqnarray}
\begin{eqnarray}
\lefteqn{\oint_{S_s}{\bf{a}}_{n}\times{\bf{h}}^{1}\,dS = \mbox{} - {\bf{a}}_{y}\times\left[{\bf{H}}^{\rmA 1}-{\bf{H}}^{\rmB 1}\right]\,{\hat{S}_p}} \nonumber \\
& & \mbox{} - {\bf{a}}_{y}\times\left[\hat{V}_{sA} \frac{\partial {\bf{H}}^{\rmA 0}}{{\partial}{\hat y}} +\hat{V}_{sB} \frac{\partial {\bf{H}}^{{B0}}}{{\partial}{\hat y}} \right]_{y=0} \label{bcscath1} \\
& & \mbox{} +\oint_{S_s} \bfxi {\bf a}_n \cdot \nabla_{\xi} \times {\bf h}^1 \, dS \,\,\, . \nonumber
\end{eqnarray}
Substituting these into (\ref{solvhm}), we get a solvability condition for $m=0$:
\begin{eqnarray}
 \lefteqn{\mathbf{a}_y \times \left[{\bf{H}}^{\rmA 0}({\bf r}_o) -{\bf{H}}^{\rmB 0}({\bf r}_o)\right] =} \label{solvhm0} \\
 && \int_V \nabla_{\xi} \times \mathbf{h}^0 \, dV + \oint_{S_s} \bfxi {\bf a}_n \cdot \nabla_{\xi} \times {\bf h}^0 \, dS \nonumber
\end{eqnarray}
and for $m=1$:
\begin{eqnarray}
 \lefteqn{\mathbf{a}_y \times \left[{\bf{H}}^{\rmA 1}({\bf r}_o) -{\bf{H}}^{\rmB 1}({\bf r}_o)\right] =} \label{solvhm1} \\
 &&  \int_V \nabla_{\xi} \times \mathbf{h}^1 \, dV + \oint_{S_s} \bfxi {\bf a}_n \cdot \nabla_{\xi} \times {\bf h}^1 \, dS \nonumber \\
 && \mbox{} - {\bf{a}}_{y}\times\left[\hat{V}_{sA} \frac{\partial {\bf{H}}^{\rmA 0}}{{\partial}{\hat y}} +\hat{V}_{sB} \frac{\partial {\bf{H}}^{{B0}}}{{\partial}{\hat y}} \right]_{y=0} \nonumber
\end{eqnarray}

\section{Other Integrals of the Zeroth-Order Boundary-Layer Fields}
\label{ap0bl}

A number of integrals of the zeroth-order boundary-layer fields over the period cell can be evaluated by appropriate use of Stokes' theorem or the divergence theorem, by methods similar to those used in Appendix~\ref{apbl}. This will allow simplification of the expressions in the main derivations. For example, by (\ref{peoa}) we can write $\nabla_{\xi} \times \left( \xi_y \mathbf{e}^0 \right) = \mathbf{a}_y \times \mathbf{e}^0$. Integrating this equation over the volume $V_A$ or $V_B$ and using the generalized Stokes theorem and relevant boundary and periodicity conditions gives

\begin{equation}
 \mathbf{a}_y \times \int_{V_{(A,B)}} \mathbf{e}^0 \, dV = - \int_{(\partial A_s, \partial B_s)} \xi_y \mathbf{a}_n \times \mathbf{e}^0 \, dS
\end{equation}
Using (\ref{peoc}), (\ref{peod}) and (\ref{Vwa}) we have finally
\begin{equation}
 \mathbf{a}_y \times \int_{V_{(A,B)}} \mathbf{e}^0 \, dV = \hat{V}_{s(A,B)} \mathbf{a}_y \times {\bf{E}}^{(\rmA , \rmB )0}({\bf r}_o)
 \label{eint0}
\end{equation}
In a similar manner, starting from the relation $\nabla_{\xi} \cdot \left( \xi_y \mu_r \mathbf{h}^0 \right) = \mu_r \mathbf{a}_y \cdot \mathbf{h}^0$ that follows from (\ref{phob}) and using the divergence theorem, we can obtain the result
\begin{equation}
 \mathbf{a}_y \cdot \int_{V_{(A,B)}} \mathbf{h}^0 \, dV = \hat{V}_{s(A,B)} H_y^{(\rmA , \rmB )0}({\bf r}_o)
 \label{hint0}
\end{equation}
Finally, by analogous techniques we also obtain the relations
\begin{equation}
\label{pyn0}
 \mathbf{a}_y \cdot \int_{V_{(A,B)}} \mathbf{e}^0 \, dV = - \int_{(\partial A_s, \partial B_s)} \xi_y \mathbf{a}_n \cdot \mathbf{e}^0 \, dS
\end{equation}
and
\begin{equation}
\label{myn0}
 \mathbf{a}_y \times \int_{V_{(A,B)}} \mathbf{h}^0 \, dV = - \int_{(\partial A_s, \partial B_s)} \xi_y \mathbf{a}_n \times \mathbf{h}^0 \, dS
\end{equation}
which are not explicit evaluations because the values for normal $\mathbf{e}^0$ and tangential $\mathbf{h}^0$ are not known \emph{a priori} on the boundary of the scatterer.

Alternative formulas for some integrals can be evaluated by integrating expressions containing $\xi_x$ or $\xi_z$ over $V_A$ or $V_B$ and using the divergence theorem as above. Many of the steps are similar, except that now the presence of $\xi_{x,z}$ in the integrand means that not all integrals over sidewall boundary pairs ($\partial A_1$ and $\partial A_2$, for example) will cancel. The details will be omitted, and we will present only the final results needed in this paper. From the volume integral of $\nabla_{\xi} \cdot (\xi_{x,z} \mathbf{d}^0)$ we get
\begin{equation}
 \mathbf{a}_{x,z} \cdot \int_{V} \mathbf{d}^0 \, dV = - \oint_{S_s} \xi_{x,z} \mathbf{a}_n \cdot \mathbf{d}^0 \, dS + \mathbf{a}_{x,z} \cdot \int_{S_{2,4}} \mathbf{d}^0 \, dS
 \label{pynx}
\end{equation}
Integration of $\nabla_{\xi} \times (\xi_{x,z} \mathbf{h}^0)$ leads to
\begin{equation}
 \mathbf{a}_{x,z} \times \int_{V} \mathbf{h}^0 \, dV = - \oint_{S_s} \xi_{x,z} \mathbf{a}_n \times \mathbf{h}^0 \, dS + \mathbf{a}_{x,z} \times \int_{S_{2,4}} \mathbf{h}^0 \, dS
 \label{mynx}
\end{equation}
Integration of $\nabla_{\xi} \times (\xi_{x,z} \mathbf{e}^0)$ gives
\begin{equation}
\begin{aligned}
 \mathbf{a}_{x,z} &\times \int_{V} \mathbf{e}^0 \, dV = \mathbf{a}_{x,z} \times \int_{S_{2,4}} \mathbf{e}^0 \, dS \\
 \mbox{}&+ \mathbf{a}_{x,z} \times \left[ \hat{V}_s \mathbf{E}_t^0(\mathbf{r}_o) + \mathbf{a}_y \left( \frac{\hat{V}_{sA}}{\epsilon_A} + \frac{\hat{V}_{sB}}{\epsilon_B} \right) \frac{D_y^0(\mathbf{r}_o)}{\epsilon_0} \right]
 \end{aligned}
 \label{eynx}
\end{equation}
and from $\nabla_{\xi} \cdot (\xi_{x,z} \mathbf{b}^0)$ we obtain
\begin{equation}
\begin{aligned}
 \mathbf{a}_{x,z} \cdot \int_{V} \mathbf{b}^0 \, dV = \mathbf{a}_{x,z} \cdot \int_{S_{2,4}} \mathbf{b}^0 \, dS \\
 \mbox{}+ \mu_0 \left( \mu_A \hat{V}_{sA} + \mu_B \hat{V}_{sB} \right) \mathbf{a}_{x,z} \cdot \mathbf{H}_t^0(\mathbf{r}_o)
 \end{aligned}
 \label{hynx}
\end{equation}

Two final relationships involving a component of the last term of (\ref{mynx}) can be obtained by integrating $\nabla_{\xi}\times(\xi_x \mathbf{h}^0)$ over the surface $S_4 = \partial A_4 \cup \partial B_4$ at $\xi_z = 1$ and using Stokes' theorem to obtain
\begin{equation}
{\bf{a}}_x \times \int_{S_4} \mathbf{h}^0 \,dS = {\bf{a}}_x \times \left. \int_{-\infty}^{\infty} \mathbf{h}^0 \,d\xi_y \right|_{\xi_x = \xi_z = 1} \,\,\, ,
\label{be10}
\end{equation}
Similarly, by integrating $\nabla_{\xi}\times(\xi_z \mathbf{h}^0)$ over the surface $S_2 = \partial A_2 \cup \partial B_2$ at $\xi_x = 1$, we obtain
\begin{equation}
{\bf{a}}_z \times \int_{S_2} \mathbf{h}^0 \,dS = {\bf{a}}_z \times \left. \int_{-\infty}^{\infty} \mathbf{h}^0 \,d\xi_y \right|_{\xi_x = \xi_z = 1} \,\,\, .
\label{be50}
\end{equation}
The $z$-component of (\ref{be10}) gives
\begin{equation}
{\bf{a}}_y \cdot \int_{S_4} \mathbf{h}^0 \,dS = {\bf{a}}_y \cdot \left. \int_{-\infty}^{\infty} \mathbf{h}^0 \,d\xi_y \right|_{\xi_x = \xi_z = 1} \,\,\, ,
\label{be20}
\end{equation}
while the $x$-component of (\ref{be50}) gives
\begin{equation}
{\bf{a}}_y\cdot\int_{S_4}\mathbf{h}^0 \,dS = {\bf{a}}_y \cdot \left. \int_{-\infty}^{\infty} \mathbf{h}^0 \,d\xi_y \right|_{\xi_x = \xi_z = 1} \,\,\, .
\label{be30}
\end{equation}
Since both line integrals are along the same path, equating (\ref{be20}) and (\ref{be30}) gives:
\begin{equation}
{\bf{a}}_y \cdot \int_{S_2} \mathbf{h}^0 \,dS = {\bf{a}}_y \cdot \int_{S_4} \mathbf{h}^0 \,dS \,\,\, .
\label{be40}
\end{equation}
An exactly similar relation holds for $\mathbf{e}^0$.

\section{Normalized Boundary-Layer Fields}
\label{apa}

All the normalized boundary-layer fields must be periodic in $\xi_x$ and $\xi_z$, and decay exponentially to zero as $\xi_y \rightarrow \pm \infty$. The subscript $i = 1, 2$ or $3$ indicates in what direction the ``source field'' is for the given normalized field; $i=1$ for $x$, $i=2$ for $y$ and $i=3$ for $z$.

From the definitions given in (\ref{dsce}) and (\ref{peo}), the $\sce_{i}$ are found to obey
\begin{subequations}
\label{sce1}
 \begin{align}
\mbox{\rm for ${\bfxi}\,\in\,\, V$:} \qquad \ \ \ \nabla_{\xi}\times{{\sce_{i}}} &=0 \label{sce1a} \\
\mbox{\rm for ${\bfxi}\,\in\,\, V$:} \qquad \nabla_{\xi}\cdot\left( \epsilon_r {{\sce_{i}}} \right) &=0 \label{sce1b} \\
 {\mathbf{a}}_{n}\times \left( {{\sce}}_{i} + \frac{1}{q_i} \mathbf{a}_i \right)_{S_s} &= 0 \label{sce1c} \\
\left.{\bf{a}}_{y}\cdot\left[\epsilon_A\sce_{i}^{\rmA}-\epsilon_B\sce_{i}^{\rmB}\right]\right|_{{\partial}A_{g}/\partial B_{g}} &=0\ \label{sce1d} \\
\left.{\bf{a}}_{y}\times\left[\sce_{i}^{\rmA}-\sce_{i}^{\rmB}\right]\right|_{{\partial}A_{g}/\partial B_{g}} &=0
  \end{align}
\end{subequations}
where
\begin{equation}
 \begin{aligned}
  q_i &= 1 \quad \mbox{\rm for $i = 1$ or $3$}; \\
   &= \epsilon_r \quad \mbox{\rm for $i = 2$}.
 \end{aligned}
 \label{qi}
\end{equation}
Similarly, from the definitions given in (\ref{dsch}) and (\ref{pho}), the $\sch_{i}$ are found to obey
\begin{subequations}
\label{scb1}
 \begin{align}
 \mbox{\rm for ${\bfxi}\,\in\,\, V$:} \qquad \quad \ \nabla_{\xi}\times{{\sch_{i}}} &=0 \label{scb1a} \\
 \mbox{\rm for ${\bfxi}\,\in\,\, V$:} \qquad \nabla_{\xi}\cdot\left(\mu_r{{\sch_{i}}}\right) &=0 \\
 {\bf{a}}_n\cdot \left( {{\sch_{i}}} + \frac{1}{r_i} \mathbf{a}_i \right)_{\partial S_s} &= 0 \\
 \left.{\mathbf{a}}_{y}\times\left[{{\sch}}^{\rmA}_{i}-{{{\sch}}^{\rmB}_{i}} \right] \right|_{{\partial}A_{g}/\partial B_{g}} &=0 \\
 \left.{\mathbf{a}}_{y}\cdot\left[\mu_A{{\sch}}^{\rmA}_{i}-\mu_B{{\sch}}^{\rmB}_{i} \right] \right|_{{\partial}A_{g}/\partial B_{g}} &=0
 \end{align}
\end{subequations}
where
\begin{equation}
 \begin{aligned}
  r_i &= 1 \quad \mbox{\rm for $i = 1$ or $3$}; \\
   &= \mu_r \quad \mbox{\rm for $i = 2$}.
 \end{aligned}
 \label{ri}
\end{equation}
We will denote the values of $q_i$ and $r_i$ in $V_{(A,B)}$ as $q_{i(A,B)}$ and $r_{i(A,B)}$ respectively.

\section{Surface Susceptibilities}
\label{aplast}

The electric surface susceptibilities are given by
\begin{equation}
\begin{array}{c}
\chi_{ES}^{y(x,z)}=-p\left[\alpha_{Ey(x,z)}^{A}+\alpha_{Ey(x,z)}^{B}\right]\\
\chi_{ES}^{yy}=-p\left[\alpha_{Eyy}^{A}+\alpha_{Eyy}^{B}-\hat{V}_{s}\right]\\
\chi_{ES}^{xx}=p\left[\epsilon_A\left(\alpha_{Exx}^{A}-\hat{V}_{sA}\right)+\epsilon_B\left(\alpha_{Exx}^{B}-\hat{V}_{sB}\right)\right]\\
\chi_{ES}^{zz}=p\left[\epsilon_A\left(\alpha_{Ezz}^{A}-\hat{V}_{sA}\right)+\epsilon_B\left(\alpha_{Ezz}^{B}-\hat{V}_{sB}\right)\right]\\
\chi_{ES}^{(x,z)y}=p\left[\epsilon_A\alpha_{E(x,z)y}^{A}+\epsilon_B\alpha_{E(x,z)y}^{B}\right]\\
\chi_{ES}^{(xz,zx)}=p\left[\epsilon_A\alpha_{E(xz,zx)}^{A}+\epsilon_B\alpha_{E(xz,zx)}^{B}\right]\,\,\, , \\
\end{array}
\label{chiEE}
\end{equation}
and the magnetic surface susceptibilities are given by
\begin{equation}
\begin{array}{c}
\chi_{MS}^{y(x,z)}=-p\left[\alpha_{My(x,z)}^{A}+\alpha_{My(x,z)}^{B}\right]\\
\chi_{MS}^{yy}=-p\left[\alpha_{Myy}^{A}+\alpha_{Myy}^{B}-\hat{V}_{s}\right]\\
\chi_{MS}^{xx}=p\left[\mu_A\left(\alpha_{Mxx}^{A}-\hat{V}_{sA}\right)+\mu_B\left(\alpha_{Mxx}^{B}-\hat{V}_{sB}\right)\right]\\
\chi_{MS}^{zz}=p\left[\mu_A\left(\alpha_{Mzz}^{A}-\hat{V}_{sA}\right)+\mu_B\left(\alpha_{Mzz}^{B}-\hat{V}_{sB}\right)\right]\\
\chi_{MS}^{(x,z)y}=p\left[\mu_A\alpha_{M(x,z)y}^{A}+\mu_B\alpha_{M(x,z)y}^{B}\right]\\
\chi_{MS}^{(xz,zx)}=p\left[\mu_A\alpha_{M(xz,zx)}^{A}+\mu_B\alpha_{M(xz,zx)}^{B}\right]\\
\end{array}
\label{chiMM}
\end{equation}
where the various terms $\alpha_E$ and $\alpha_M$ are defined as
\begin{equation}
\begin{array}{c}
\alpha^{(A,B)}_{Ey(x,y,z)}={\bf{a}}_y\cdot \int_{V_{(A,B)}} \sce_{(1,2,3)} dV_{\xi} \\
\alpha^{(A,B)}_{Mx(x,y,z)}={\bf{a}}_x\cdot \int_{V_{(A,B)}} \sch_{(1,2,3)} dV_{\xi} \\
\alpha^{(A,B)}_{Mz(x,y,z)}={\bf{a}}_z\cdot \int_{V_{(A,B)}} \sch_{(1,2,3)} dV_{\xi} \\
\end{array}\,\,\,,
\label{definealpha}
\end{equation}
\begin{equation}
\begin{array}{c}
\alpha^{(A,B)}_{My(x,y,z)}={\bf{a}}_y\cdot \int_{S_2(A,B)} \sch_{(1,2,3)} dS_2(A,B) \\
\alpha^{(A,B)}_{Ex(x,y,z)}={\bf{a}}_x\cdot \int_{S_2(A,B)} \sce_{(1,2,3)} dS_2(A,B) \\
\alpha^{(A,B)}_{Ez(x,y,z)}={\bf{a}}_z\cdot \int_{S_4(A,B)} \sce_{(1,2,3)} dS_4(A,B) \\
\end{array}\,\,\, ,
\label{definealpha2}
\end{equation}
where the planes $S_{2A}$ and $S_{4B}$ correspond to the portions of $S_2$ in regions $A$ and $B$, respectively, and $S_{4A}$ and $S_{4B}$ correspond to the portions of $S_4$ in region $A$ and $B$, respectively. The subscripts and superscripts in these parameters have the following meanings. The superscript $(A, B)$ corresponds to an integral over either $V_A$ or $V_B$.  The first subscript ($E$ or $M$) indicates an integral of either an $\sce$-field or a $\sch$-field.  The second subscript corresponds to the $x$ or $y$ component of $\bfalpha_{E,M}$. The third subscript corresponds to the component of the excitation field that generates $\sce_i$ or $\sch_i$.

In deriving these surface susceptibilities we used a procedures similar to those in Appendix~C of \cite{wirehk} to show that $\int\sce_2\, dV$ has only a $y$-component while $\int\sch_{1,3}\, dV$ have no $y$-components, so that some of the integrals of the fields can be simplified as
\begin{equation}
\begin{array}{rcl}
\int_{AB} \sce_1 dV_{\xi}&=& {\bf{a}}_x V_{s}+{\bf{a}}_y\left[ \alpha^{A}_{Eyx}+\alpha^{B}_{Eyx}\right] \\
\int_{AB} \sce_2 dV_{\xi}&=& {\bf{a}}_y\left[ \alpha^{A}_{Eyy}+ \alpha^{B}_{Eyy}\right] \\
\int_{AB} \sce_3 dV_{\xi}&=& {\bf{a}}_z V_{s}+{\bf{a}}_y \left[\alpha^{A}_{Eyz}+\alpha^{B}_{Eyz}\right] \\
\end{array}
\label{eq2}
\end{equation}
\begin{equation}
\begin{array}{c}
\int_{(A,B)} \sch_1 dV_{\xi}= {\bf{a}}_x \alpha^{(A,B)}_{Mxx}+{\bf{a}}_z \alpha^{(A,B)}_{Mzx}\\
\int_{(A,B)} \sch_2 dV_{\xi}= {\bf{a}}_y V_{s(A,B)}+{\bf{a}}_x \alpha^{(A,B)}_{Mxy}+{\bf{a}}_z \alpha^{(A,B)}_{Mzy} \\
\int_{(A,B)} \sch_3 dV_{\xi}= {\bf{a}}_x \alpha^{(A,B)}_{Mxz}+{\bf{a}}_z \alpha^{(A,B)}_{Mzz}  \,\,\, . \\
\end{array}
\label{eq3}
\end{equation}


\begin{thebibliography}{999}

\bibitem{kmh} E. F. Kuester, M. A. Mohamed, M. Piket-May and C.L. Holloway,
``Averaged transition conditions for electromagnetic fields at a
metafilm,'' {\it IEEE Trans. Ant. Prop.}, vol. 51,
pp. 2641-2651, 2003.

\bibitem{c1} {\it Advances in Electromagnetics of Complex Media and
Metamaterials} (S. Zouhdi, A. Sihvola, and M. Arsalane, eds.).
Kluwer Academic Pub.: Boston, 2002.

\bibitem{c1c} C. Caloz and T. Itoh, {\it Electromagnetic Metamaterials : Transmission Line
Theory and Microwave Applications}. Wiley-IEEE Press: 2005.

\bibitem{c1ca} G.V. Eleftheriades and K.G. Balmain, {\it Negative Refraction
Metamaterials: Fundamental Principles and Applications}. Wiley,
2005.

\bibitem{c1b} N. Engheta and R.W. Ziolkowski, {\it Electromagnetic Metamaterials: Physics
and Engineering Explorations}. John Wiley \& Sons: 2006.

\bibitem{marq} R. Marqu\'{e}s, F. Mart\'{i}n and M. Sorolla, {\it Metamaterials with Negative Parameters: Theory, Design, and Microwave Applications}. Hoboken, NJ: Wiley-Interscience, 2008.

\bibitem{cap} F. Capolino (ed.), {\it Metamaterials Handbook: Theory and Phenomena of Metamaterials}. Boca Raton, FL: CRC Press, 2009.

\bibitem{cui} T. J. Cui, D. R. Smith and R. Liu (eds.), {\it Metamaterials: Theory, Design, and Applications}. New York: Springer, 2010.


\bibitem{hk3} C. L. Holloway, E. F. Kuester, J. A. Gordon, J. O'Hara, J. Booth and D. R. Smith, ``An overview of the theory and applications of metasurfaces: The two-dimensional equivalents of metamaterials'', {\it IEEE Ant. Prop. Mag.}, vol. 54, no. 2, pp. 10-35, April 2012.

\bibitem{alex} A.A. Maradudin, ed., {\it Structured Surfaces as Optical Metamaterials}. Cambridge University Press: Cambridge, UK, 2011.


\bibitem{surfacewave} C.L. Holloway, D.C. Love, E. F. Kuester,J.A. Gordon, and D.A. Hill, ``Use of generalized sheet transition conditions to model guided waves on metasurfaces/metafilms'', {\it IEEE Trans. Ant. Prop.}, vol. 60, pp. 5173-5186, 2012.

\bibitem{wirehk} C.L. Holloway, E.F. Kuester, ``A homogenization technique for obtaining generalized sheet transition conditions
for an arbitrarily shaped coated wire grating,'' {\it Radio Science}, vol. 49, no. 10, pp. 813-7850, 2014.


\bibitem{hk2} C.L. Holloway, A. Dienstfrey, E.F. Kuester, J.F. O'Hara,
A.K. Azad and A.J. Taylor, ``A discussion on the interpretation and
characterization of metafilms-metasurfaces: The two-dimensional
equivalent of metamaterials'', {\it Metamaterials}, vol. 3, pp. 100-112, 2009.

\bibitem{awpl} C.L. Holloway, E.F. Kuester and A. Dienstfrey, ''Characterizing metasurfaces/metafilms: The connection between surface susceptibilities and effective material properties'', {\it IEEE Ant. Wireless Prop. Lett.}, vol. 10, pp. 1507-1511, 2011.


\bibitem{ss1} {A. Stahl and H. Wolters, ``Elektromagnetische randbedingungen und oberfl\"{a}cheneffekte in
ph\"{a}nomenologischer Sicht,'' {\it Z. Physik}, vol. 255, pp. 227-239, 1972.}

\bibitem{ss2} {P. Guyot-Sionnest, W. Chen, and Y.R. Shen, ``General
considerations on optical second-harmonic generation from surface
and interfaces,'' {\it Phys. Rev. B}, vol. 33, pp. 8254-8263, 1986.}

\bibitem{ss3} {R. Atkinson and N.F. Kubrakov, ``Magneto-optical
characterization of ferromagnetic ultrathin multilayers in terms
of surface susceptibility tensors,'' {\it Phys. Rev. B}, vol. 66, art. 024414, 2002.}

\bibitem{calzo} Dimitriadis, A.~I., D.~L. Sounas, N.~V. Kantartzis, C. Caloz, and T.~D. Tsiboukis, ``Surface susceptibility bianisotropic matrix model for periodic metasurfaces of uniaxially mono-anisotropic scatterers under oblique TE-wave incidence,'' {\it IEEE Trans. Ant. Prop.}, vol. 60, 5753-5767, 2012.


\bibitem{hr1} E. Sanchez-Palencia, ``Comportements local et macroscopique d'un type de milieux physique h\'{e}t\'{e}rogenes,'' {\it Int. J. Eng. Sci.}, vol. 12, pp. 331-351, 1974.

\bibitem{hr4} E.F. Kuester and C. L. Holloway, ``A low-frequency model for wedge or pyramid absorber arrays--I: Theory,'' {\em IEEE Trans. Electromag. Compat.,} vol. 36, pp. 300-306, 1994.

\bibitem{hr2} E. Sanchez-Palencia, {\it Non-Homogeneous Media and Vibration Theory} (Lecture Notes in Physics no. 127). Berlin: Springer-Verlag, 1980, pp. 68-77.
\bibitem{hr3} A. Bensoussan, J.-L. Lions and G. Papanicolaou, {\it Asymptotic Analysis for Periodic Structures}. Amsterdam: North-Holland, 1978.

\bibitem{hr3b} N. Bakhvalov and G. Panasenko, {\it Homogenisation: Averaging Processes in Periodic Media}. Dordecht: Kluwer Academic Publishers, 1989.

\bibitem{hr5} C.L. Holloway, and E. F. Kuester, ``Impedance-type boundary conditions for a periodic interface between a dielectric and highly conducting medium,'' {\it IEEE Trans. Ant. Prop.}, vol. 48, pp. 1660-1672, 2000.

\bibitem{hr6} C. L. Holloway and E. F. Kuester, ``Equivalent boundary conditions for a perfectly conducting periodic surface with a cover layer,'' {\it Radio Sci.}, vol. 35, pp. 661-681, 2000.

\bibitem{hr7} C.L. Holloway, and E.~F. Kuester, ``Corrections to the classical continuity boundary conditions at the interface of a composite medium,`` {\it Photonics and Nanostructures: Fundamentals and Applications}, vol. 11, pp. 397-422, 2013.

\bibitem{delourme1}  B. Delourme,  ``On the well-posedness, stability and accuracy of an asymptotic model for thin periodic interfaces in electromagnetic scattering problems,'' {\it Mathematical Models and Methods in Applied Sciences}, vol. 23, pp. 2433-2464, 2013.

\bibitem{delourme2}  B. Delourme, ``High-order asymptotics for the electromagnetic scattering by thin periodic layers,'' {\it Mathematical Methods in the Applied Sciences}, vol. 38, pp. 811-833, 2015.

\bibitem{vain} L.A. Wainstein, ``On the electrodynamic theory of grids'', in {\it Elektronika Bol'shikh Moshchnostei}, vol. 2 (P.~L. Kapitza and L.~A. Wainstein, editors). Moscow: Nauka, pp. 26-74, 1963 [in Russian; Engl. transl. in {\it High-Power Electronics}, vol. 2.  Oxford: Pergamon Press, 1966, chapter II, pp. 14-48].

\bibitem{senior}  T.~B.~A. Senior, and J.~L. Volakis, {\it Approximate Boundary Conditions in Electromagnetics}. London: Institution of Electrical Engineers, p. 163, 1995.

\bibitem{pere1} M.~L. Pereyaslavets, ``Relationship between tensors in the joining boundary conditions at semitransparent surface,'' {\it Radiotekh. Elektron.}, vol. 37, pp. 1559-1564, 1992 [in Russian; Engl. transl. in {\it J. Commun. Technol. Electron.}, vol. 38, no. 1, pp. 23-29].

\bibitem{pere2} M.~L. Pereyaslavets, ``The reciprocity and power conservation principles for boundary conditions on a semitransparent surface,'' {\it IEEE Trans. Ant. Prop.}, vol. 42, pp. 449-452, 1994.

\bibitem{duru} M. Durufl\'{e}, V. P\'{e}ron and C. Poignard, ``Thin layer models for electromagnetism,'' \emph{Commun. Comput. Phys.}, vol. 16, pp. 213-238, 2014.

\bibitem{chun} S. Chun, H. Haddar and J.S. Hesthaven, ``High-order accurate thin layer approximations for time-domain electromagnetics, Part II: Transmission layers,'' \emph{J. Comp. Appl. Math.}, vol. 234, pp. 2587-2608, 2010.

\bibitem{keller} J. B. Keller, ``Conductivity of a medium containing a dense array of perfectly conducting spheres or cylinders or nonconducting cylinders,'' {\it J. Appl. Phys.}, vol. 34, pp. 991-993, 1963.

\bibitem{sangani} A. S. Sangani and A. Acrivos, ``The effective conductivity of a periodic array of spheres,'' {\it Proc. Roy. Soc. London A}, vol. 386, pp. 263-275, 1983.

\bibitem{andrian} I. Andrianov, V. Danishevskyi and S. Tokarzewski, ``Two-point quasifractional approximants for effective conductivity of a simple cubic lattice of spheres,'' {\it Int. J. Heat Mass Transfer}, vol. 39, pp. 2349-2352, 1996.

\bibitem{powell} D.A. Powell and Y.S. Kivshar, ``Substrate-induced bianisotropy in metamaterials,'' \emph{Appl. Phys. Lett.}, vol. 97, art. 091106, 2010.

\bibitem{alboo} M. Albooyeh and C.R. Simovski, ``Substrate-induced bianisotropy in plasmonic grids,'' \emph{J. Opt.}, vol. 13, art. 105102, 2011.

\bibitem{alboo2} M. Albooyeh, D. Morits and C.R. Simovski, ``Electromagnetic characterization of substrated metasurfaces,'' \emph{Metamaterials}, vol. 5, pp. 178-205, 2011.

\bibitem{bouch1} G. Bouchitt\'{e} and D. Felbacq, ``Homogenization near resonances and artificial magnetism from dielectrics,'' {\it Comptes Rendus Acad. Sci. Paris, ser. I}, vol. 339, pp. 377-382, 2004.

\bibitem{bouch2} G. Bouchitt\'{e}, C. Bourel and D. Felbacq, ``Homogenization of the 3D Maxwell system near resonances and artificial magnetism,'' {\it Comptes Rendus Acad. Sci. Paris, ser. I}, vol. 347, pp. 571-576, 2009.

\bibitem{bouch3} D. Felbacq, B. Guizal, G. Bouchitt\'{e} and C. Bourel, ``Resonant homogenization of a dielectric metamaterial,'' {\it Micr. Opt. Technol. Lett.}, vol. 51, pp. 2695-2701, 2009.

\bibitem{bouch4} G. Bouchitt\'{e} and B. Schweizer, ``Homogenization of Maxwell’s equations in a split ring geometry,'' {\it Multiscale Model. Simul.}, vol. 8, pp. 717-750, 2010.


\end{thebibliography}
\end{document}